\documentclass[journal]{IEEEtran}
\usepackage{amsmath,amsfonts}
\usepackage{amssymb}
\usepackage{algorithm}
\usepackage{algorithmic}
\usepackage{array}
\usepackage[caption=false,font=footnotesize,labelfont=rm,textfont=rm]{subfig}
\usepackage{textcomp}
\usepackage{stfloats}
\usepackage{url}
\usepackage{verbatim}
\usepackage{graphicx}
\usepackage{cite}
\usepackage{xcolor}
\usepackage[colorlinks=true, linkcolor=blue, citecolor=blue, urlcolor=blue]{hyperref}
\usepackage{amsthm}
\usepackage{booktabs}  
\usepackage{multirow}  
\usepackage{makecell}

\usepackage{float}
\newtheorem{theorem}{Theorem}

\begin{document}

\title{Energy-Workload Coupled Migration Optimization Strategy for Virtual Power Plants with Data Centers Considering Fuzzy Chance Constraints}

\author{Jia-Kai Wu, {\em{Graduate Student Member, IEEE}},
        Zhi-Wei Liu, {\em{Senior Member, IEEE}},
        Yong Zhao,
        Yan-Wu Wang, {\em{Senior Member, IEEE}},
        Fan-Rong Qu,
        and Chaojie Li, {\em{Member, IEEE}}%
        \thanks{Jia-Kai Wu, Zhi-Wei Liu, Yong Zhao, Yan-Wu Wang, and Fan-Rong Qu are with the School of Artificial Intelligence and Automation, Huazhong University of Science and Technology, Wuhan 430074, China (e-mail: jkwu@hust.edu.cn; zwliu@hust.edu.cn; zhiwei98530@hust.edu.cn; wangyw@hust.edu.cn; frqu@hust.edu.cn).
        
       Chaojie Li is with the Department of Electrical Engineering, City University of Hong Kong, Hong Kong, China (e-mail: chaojili@cityu.edu.hk).}%
}

\maketitle

\begin{abstract}
This paper proposes an energy-workload coupled migration optimization strategy for virtual power plants (VPPs) with data centers (DCs) to enhance resource scheduling flexibility and achieve precise demand response (DR) curve tracking. A game-based coupled migration framework characterized by antisymmetric matrices is first established to facilitate the coordination of cross-regional resource allocation between VPPs. To address the challenge posed to conventional probabilistic modeling by the inherent data sparsity of DC workloads, deterministic equivalent transformations of fuzzy chance constraints are derived based on fuzzy set theory, and non-convex stochastic problems are transformed into a solvable second-order cone program. To address the multi-player interest coordination problem in cooperative games, an improved Shapley value profit allocation method with the VPP operator as intermediary is proposed to achieve a balance between theoretical fairness and computational feasibility. In addition, the alternating direction method of multipliers with consensus-based variable splitting is introduced to solve the high-dimensional non-convex optimization problem, transforming coupled antisymmetric constraints into separable subproblems with analytical solutions. Simulations based on real data from Google's multiple DCs demonstrate the effectiveness of the proposed method in improving DR curve tracking precision and reducing operational costs.
\end{abstract}


\section{Introduction}
\IEEEPARstart{W}{ith} the rapid development of the global digital economy, particularly the widespread adoption of large-scale artificial intelligence model training, the annual electricity consumption of global data centers (DCs) has reached 800 terawatt-hours \cite{Statista2024DataCenters,liu2023placement,zhang2022two}.  
It is projected that the electricity consumption of global data DCs will more than double in the next five years, with a compound annual growth rate of 19.5\%  \cite{IDC2024AISemiconductor}.
This trend not only places significant pressure on energy supply but also presents a crucial challenge to carbon emission reduction targets, leading to extensive attention from both academia and industry on energy optimization for DCs.

The functionality of large-scale DC clusters in power systems extends beyond that of traditional load nodes, exhibiting temporal and spatial coupled-dimensional regulation characteristics \cite{chen2020internet}. In the temporal domain, these systems achieve power regulation responses within microsecond to millisecond timeframes through refined server load management and dynamic task scheduling \cite{li2023computation}. In the spatial domain, leveraging inter-regional computational migration mechanisms among geographically distributed DCs realizes effective virtual power transfer \cite{wu2023incentivizing,kozyrakis2013resource,ruan2024privacy}. This composite network of intertwined energy and information flows enables data centers to exhibit response characteristics markedly distinct from traditional controllable loads \cite{zhang2022peer,zheng2024premium,kang2021novel,zhang2023equilibrium}, introducing greater flexibility and complexity to power system operations and optimization.

The coupled regulation capability and diversified energy supply structure of DCs are highly compatible with the customer directrix load (CDL) demand response (DR) mechanism \cite{meng2023transmission,wang2024incentive}.  
Unlike traditional DR, the CDL mechanism requires precise full-time load tracking rather than simple peak shaving and valley filling. 
This aligns perfectly with three key technical characteristics of DCs: 1) spatial-temporal coupled-dimensional regulation capability providing greater load adjustment range; 2) categorizable and delayable computational tasks meeting the temporal flexibility requirements of CDL tracking; and 3) diversified energy systems offering additional regulation degrees of freedom needed for precise CDL tracking. 
This deep technical compatibility makes DCs ideal carriers for implementing CDL-based DR.

Although CDL-based DR enhances system regulation accuracy, it imposes stringent tracking requirements that curtail the operational flexibility of individual participants \cite{hong2024customer,ma2024optimal}. For a single DC, this operational rigidity becomes particularly acute when confronted with stochastic renewable generation and workload arrivals, potentially degrading its core service performance \cite{meng2023transmission}. The virtual power plant (VPP) framework is introduced as a systematic approach to resolve this conflict by aggregating multiple DCs into a unified, dispatchable portfolio \cite{zhang2022optimal,wu2024coordinated,chen2021customized,xu2022competitive}. Within this cooperative construct, the control objective shifts from ensuring individual compliance to optimizing the aggregate response to meet the CDL trajectory \cite{yi2020bi}. 
This paradigm of centralized management with distributed execution leverages resource complementarity through cooperative strategies such as inter-regional energy sharing and workload migration, thereby substantially broadening the optimization space.

Relevant scholarship has approached these issues from three primary perspectives. First, at the local resource management level, optimization theories such as mixed-integer programming and model predictive control are employed for the fine-grained regulation of resources like computing, cooling, and energy storage. The objective is to achieve an optimal trade-off between internal energy efficiency and operational costs \cite{zhang2022two,long2023collaborative}. Second, at the networked coordination level, methods like distributed optimization and game theory are used to minimize total system costs through cross-regional workload migration and computing power scheduling, leveraging regional differences in electricity prices or the complementarity of renewable energy \cite{ruan2024privacy}. Third, at the energy and market level, data centers are positioned as active flexibility resources. Here, methods such as bi-level programming and market equilibrium models are used to model their grid interaction, enabling deep participation in market mechanisms like demand response and ancillary services to support stable, low-carbon grid operation \cite{koraki2017wind,cupelli2018data,zhang2022framework}.

Despite the significant progress made by these studies at different levels, several key challenges remain unaddressed. First, the coupled optimization framework for energy and workload is underdeveloped \cite{yan2024low}. Most studies still treat them in a decoupled manner or through simple linear superposition, which fails to capture their inherent coupling characteristics, thereby limiting scheduling flexibility and the deep exploration of cross-regional resource complementarity. Second, traditional uncertainty optimization methods are ill-suited for the complex environment of VPP-DC systems \cite{zhu2024multi,chen2024feasible}. The inherent data sparsity of DC workloads, particularly from emerging applications like large-scale AI model training with no historical precedent, cannot support the construction of reliable probability density functions (PDFs). While robust optimization does not require PDFs \cite{fang2022data}, its approach of guaranteeing worst-case security at the expense of economic efficiency can lead to a surge in operational costs. Finally, existing benefit coordination mechanisms face significant limitations in high-dimensional VPP environments \cite{wu2022modified}. While methods like the Shapley value, Core allocation, non-cooperative games, and bargaining strategies offer theoretical frameworks, they are often impractical for large-scale applications due to issues such as exponential computational complexity, a lack of unique solutions, poor scalability, and high sensitivity to initial parameters.

To overcome the aforementioned limitations, this paper proposes a framework where local DCs and distributed photovoltaic systems form a VPP to enable energy sharing and workload migration, and it addresses system uncertainty based on a fuzzy chance-constrained method. This method does not require PDF data and replaces the rigid worst-case paradigm of robust optimization with an adjustable risk management mechanism. By introducing a credibility measure, it allows the VPP operator to explicitly define the required confidence level, thus enabling a quantifiable and flexible trade-off between operational costs and system reliability. To address the challenge of multi-agent benefit coordination, an improved Shapley value profit allocation method is also proposed to ensure fairness and computational feasibility in profit distribution.
The main contributions of this paper are summarized as follows:
\begin{enumerate}
\item Different from existing works \cite{ruan2024privacy,zhang2022framework,yan2024low} that separate energy and workload optimization, this paper establishes a coupled migration framework using antisymmetric matrices for energy transfer and workload migration in CDL-based DR. This approach expands the optimization space by leveraging geographic complementarities, enabling more precise CDL curve tracking while ensuring system conservation via antisymmetric constraints.
\item Unlike the overly conservative robust optimization or computationally burdensome scenario-based methods \cite{fang2022data,zhu2024multi}, this paper derives the deterministic equivalent form of fuzzy chance constraints to address a type of uncertainty that is difficult for traditional methods to handle. This difficulty arises from characteristics such as data sparsity and the challenge of constructing precise probabilistic models. The proposed approach transforms the complex, non-convex stochastic problem into a tractable second-order cone programming (SOCP) problem and allows decision-makers to flexibly trade off between operational costs and system reliability via an adjustable confidence level.
\item Compared to traditional Shapley value methods \cite{wu2024coordinated,zhang2022framework} with exponential computational complexity, an improved Shapley value profit allocation framework with the VPPO as intermediary is introduced, reducing complexity from $O(2^N)$ to $O(N)$ while balancing theoretical fairness and practicality. Additionally, a distributed alternating direction method of multipliers (ADMM) algorithm with variable splitting is designed to solve high-dimensional non-convex problems.
\end{enumerate}

The rest of this paper is organized as follows: 
Section II introduces the system model and problem formulation. 
Section III details the bidirectional migration strategies and uncertainty model construction. 
Section IV presents the solution methods. 
Section V provides simulation experiments and result analysis, and Section VI concludes the study.

\section{System Model and Problem Formulation}
\subsection{System Description}
\begin{figure}[!t]
\centering
\includegraphics[width=8.6cm]{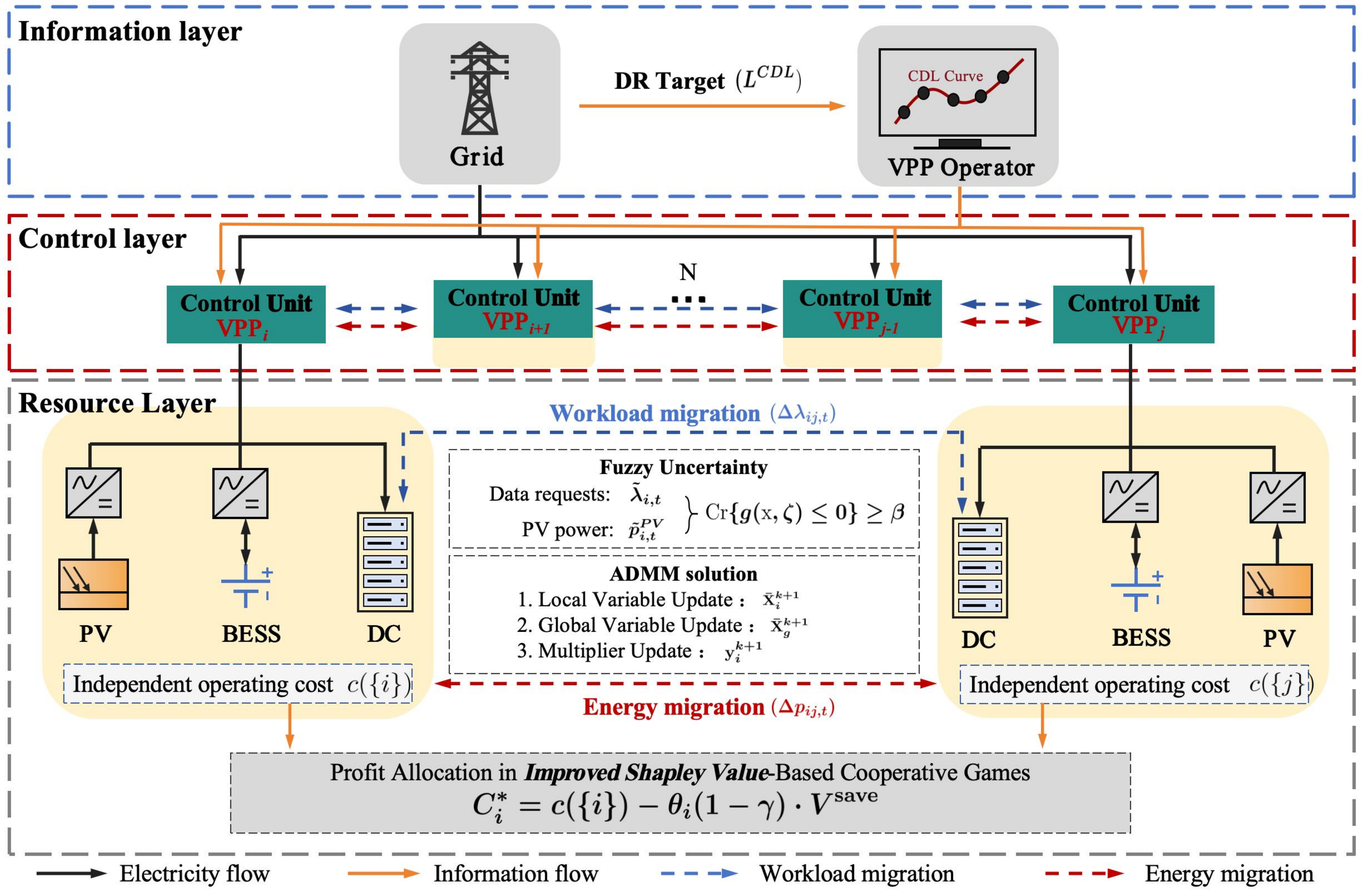}
\caption{Framework for geo-distributed DCs participating in DR}
\label{fig_framework}
\end{figure}

This paper proposes an energy-workload coupled migration framework to address CDL-based DR challenges in trajectory tracking under limited flexibility and multi-source uncertainties. 
As illustrated in Fig.~\ref{fig_framework}, the framework comprises $N$ geo-distributed VPPs (set $\mathcal{N} = \{1,2,\dots,i,\dots,N\}$) with operations discretized into scheduling periods $\mathcal{T} = \{1,2,\dots,t,\dots,T\}$. Each VPP$_i$ ($i\in \mathcal{N} $) independently manages its own local resources, which include DCs with power demand, battery energy storage systems (BESS) offering temporal energy regulation capabilities, and PV systems supplying renewable energy.  The geo-distribution enables workload migration to VPPs with lower costs or excess capacity and energy transfers to VPPs with deficits or higher costs, all orchestrated through a hierarchical three-layer coordination architecture.

The information layer forms the top tier, where the power grid communicates standardized CDL trajectories ($\mathbf{L}^{\mathrm{CDL}} $) to the VPPO. The CDL mechanism aims to minimize the Euclidean distance $\|\mathbf{L} - \mathbf{L}^{\mathrm{CDL}} \|_2$ between actual and target load shapes, where $\mathbf{L}$ represents the actual load shape. The VPPO acts as a market intermediary, coordinating the operations of multiple independent VPPs while maintaining individual interest equilibrium.

The control layer implements decision-making across the separate VPPs. Each VPP$_i$ control unit manages its own local resources while coordinating with other VPP$_j$ ($j\in \mathcal{N}$) units for energy and workload exchange. Key local decision variables for each VPP$_i$ include server scheduling vectors ($\boldsymbol{s}_i\in\mathbb{Z}^T$), batch workload energy consumption vector ($\boldsymbol{e}_i^\mathrm{b}\in\mathbb{R}_+^T$), energy storage operation matrices ($\boldsymbol{q}_i \in \mathbb{R}^{T \times 2}$), and grid interaction vectors ($\boldsymbol{p}_i^\mathrm{b} \in \mathbb{R}_+^T$). Inter-VPP coordination occurs through bidirectional transfers where $\Delta\lambda_{ij,t}$ represents interactive workload transferred from VPP$_i$ to VPP$_j$ in time slot $t$, while $\Delta p_{ij,t}$ denotes corresponding energy transfer. To maintain system-wide resource conservation, these transfer variables satisfy antisymmetric constraints where $\Delta\lambda_{ij,t}+\Delta\lambda_{ji,t}=0$ and $\Delta p_{ij,t}+\Delta p_{ji,t}=0$ for all $i,j\in \mathcal{N}$. These constraints ensure zero net transfer across the VPP network.

The resource layer forms the physical infrastructure foundation. In this layer, each VPP's DC power demands are met through a combination of its local PV generation, BESS discharge, and grid purchases when needed. The geographic distribution of these resources allows VPPs to leverage regional complementarities through coordinated operation.

The system operates in four stages: (1) VPPO receives CDL signals and analyzes VPP resource characteristics; (2) each VPP computes its characteristic function $c(\{i\})$; (3) optimization occurs using bidirectional migration mechanisms; and (4) improved Shapley value methods distribute benefits fairly among participating VPPs.

\subsection{Basic Mathematical Bodel of Each Component}
\subsubsection{DC Model}
Batch workloads are scheduled tasks with lower latency sensitivity that can be shifted in the time domain. They use dedicated servers separate from interactive workload servers $\mathbf{s}_i$. In this paper, servers and task arrival rates refer specifically to interactive workloads, while batch workloads are simplified as consuming power $e_{i,t}^b$ in DC$_i$.
Therefore, DC$_i$'s power consumption in time slot $t$ is:
\begin{align}
\label{eq:data_center_power}
p_{i,t}^{\mathrm{d}} &= s_{i,t}\left[e^{\mathrm{idle}} + \left(e^{\mathrm{peak}} - e^{\mathrm{idle}}\right)U_{i,t} + (\eta - 1)e^{\mathrm{peak}}\right]+e_{i,t}^b,
\end{align}
\begin{align}
\label{eq:server_utilization}
U_{i,t} &= \frac{\lambda_{i,t}}{s_{i,t} u_i},
\end{align}
where $s_{i,t}$ is active server count, $e^{\mathrm{idle}}$/$e^{\mathrm{peak}}$ are server idle/peak power, $\eta$ is the power usage effectiveness of the DC, $u_i$ is server processing rate, $\lambda_{i,t}$ is task arrival amount, and $e_{i,t}^b$ is batch workload energy. Denote $\hat{e}_{i}^b$ as the total energy requirement for batch workloads of VPP$_i$ over the scheduling horizon. The batch workloads must satisfy $\sum_{t \in \mathcal{T}} e_{i,t}^b = \hat{e}_{i}^b$.

Quality of service (QoS) is ensured through queuing delay cost and server capacity constraints:
\begin{align}
\label{eq:qos_cost}
C_{i}^{\mathrm{QoS}} &= \sum_{t \in \mathcal{T}} \kappa \lambda_{i,t} \left(\frac{1}{s_{i,t} u_i - \lambda_{i,t}}\right),
\end{align}
\begin{align}
\label{eq:server_constraint}
\begin{cases}
s_{i,t} \leq s^{\mathrm{max}}, \\
s_{i,t} u_i \geq \lambda_{i,t},
\end{cases}
\quad \forall t,
\end{align}
where $\kappa$ is delay cost coefficient, $s^{\mathrm{max}}$ is maximum server count, and $\frac{1}{s_{i,t}u_i - \lambda_{i,t}}$ is average response time.

\subsubsection{PV and BESS Models}
PV generation and BESS degradation costs are modeled as:
\begin{align}
C_{i}^{\mathrm{PV}} &= \sum_{t \in \mathcal{T}} \rho^{\mathrm{PV}} p_{i,t}^{\mathrm{PV}},\\
C_{i}^{\mathrm{BESS}} &= \sum_{t \in \mathcal{T}} \varepsilon \left(q_{i,t}^{\mathrm{ch}} + q_{i,t}^{\mathrm{dis}}\right),
\end{align}
\begin{align}
\label{eq:SoC}
\mathrm{SoC}_{i,t} &= (1-\epsilon)\,\mathrm{SoC}_{i,t-1} + \frac{\eta^{\mathrm{ch}}\,q_{i,t}^{\mathrm{ch}}}{B_i} - \frac{q_{i,t}^{\mathrm{dis}}}{\eta^{\mathrm{dis}}\,B_i},
\end{align}
\begin{align}
\label{eq:SoC_constraints}
\begin{cases}
\mathrm{SoC}_{i}^{\mathrm{min}} \leq \mathrm{SoC}_{i,t} \leq \mathrm{SoC}_{i}^{\mathrm{max}}, \\
0 \leq q_{i,t}^{\mathrm{ch}} \leq q_i^{\mathrm{ch,max}},  \\
0 \leq q_{i,t}^{\mathrm{dis}} \leq q_i^{\mathrm{dis,max}}, \\
\mathrm{SoC}_{i,0} = \mathrm{SoC}_{i,T}, \\
q_{i,t}^{\mathrm{ch}} \cdot q_{i,t}^{\mathrm{dis}} = 0, 
\end{cases}
\quad \forall t,
\end{align}
where $\rho^{\mathrm{PV}}$ is the PV unit cost and $p_{i,t}^{\mathrm{PV}}$ is the PV power output. For the BESS, $\varepsilon$ is the degradation coefficient; $q_{i,t}^{\mathrm{ch}}$ and $q_{i,t}^{\mathrm{dis}}$ are the charging and discharging power. $\mathrm{SoC}_{i,t}$ is the state of charge, $\epsilon$ is the self-discharge coefficient, $B_i$ is the storage capacity, and $\eta^{\mathrm{ch}}$ and $\eta^{\mathrm{dis}}$ are the charging and discharging efficiencies. The parameters $\mathrm{SoC}_{i}^{\mathrm{min}}$, $\mathrm{SoC}_{i}^{\mathrm{max}}$, $q_i^{\mathrm{ch,max}}$, and $q_i^{\mathrm{dis,max}}$ denote the minimum and maximum limits for SoC and power, respectively.

\subsection{CDL-based DR Mechanism}
The total external load (adjustable energy) defined as the grid-purchased power $p_{i,t}^{\mathrm{b}}$ must satisfy the following constraint:
\begin{align}
\label{eq:independent_external_load}
p_{i,t}^{\mathrm{b}} \geq p_{i,t}^{\mathrm{d}} - \left(p_{i,t}^{\mathrm{PV}} - q_{i,t}^{\mathrm{ch}} + q_{i,t}^{\mathrm{dis}}\right), \quad \forall t.
\end{align}

Normalize $p_{i,t}^{\mathrm{b}}$ to obtain the load curve $L_{i,t}$ as shown in \eqref{eq:CDL_constraint}:
\begin{align}
\label{eq:CDL_constraint}
\begin{cases}
L_{i,t} = \dfrac{p_{i,t}^{\mathrm{b}}}{P_i^D}, \\[10pt]
d_i = \sqrt{\sum_{t \in \mathcal{T}} \left(L_{i,t} - L_t^{\mathrm{CDL}}\right)^2},
\end{cases}
\end{align}
where $\sum_{t \in \mathcal{T}} L_{i,t} = 1$, $P_i^D$ is the declared adjustable load capacity of VPP$_i$, and $L_t^{\mathrm{CDL}}$ is the published load guideline. The Euclidean distance between the actual and baseline load curves is denoted as $d_i$, with similarity $\varepsilon_i = 1 - d_i$. The DR incentive is then calculated as:
\begin{align}
\label{eq:ind_DR}
R_i^{\mathrm{DR}} = \rho_o \, \varepsilon_i P_i^D,
\end{align}
where $\rho_o$ is the DR incentive price. This nonlinear problem can be relaxed into SOCP form (details in Section IV).

\section{Bidirectional Migration Strategies and Uncertainty Model Construction}
\subsection{Deterministic Bidirectional Migration Model}
\subsubsection{Independent Operation Costs}
Under non-cooperative scenarios, the cost minimization problem for independently operating VPPs is formulated as \textbf{(P1)}:
\begin{align}
\label{eq:independent_cost}
\min_{\mathbf{X}} \quad 
C_{i} - R_i^{\mathrm{DR}},
\end{align}
where $\mathbf{X}=\{s_{i,t}, p_{i,t}^{\mathrm{b}}, q_{i,t}^{\mathrm{ch}}, q_{i,t}^{\mathrm{dis}},e_{i,t}^b\}$ represents the set of decision variables for the independent operations of VPP$_i$. The total cost function $C_i$ is defined as:
\begin{align}
\label{eq:total_cost}
C_{i} = C_{i}^{\mathrm{PV}} + C_{i}^{\mathrm{BESS}} + \sum_{t \in \mathcal{T}} \rho_{i,t}^{\mathrm{b}} p_{i,t}^{\mathrm{b}} + C_{i}^{\mathrm{QoS}},
\end{align}
where $\rho_{i,t} ^{\mathrm{b}}$ is the price at which VPP$_i$ purchases electricity from the grid.
\subsubsection{Coupled Migration Strategy Based on Cooperative Game}
The cooperative game comprises players $i\in \mathcal{N}$ representing VPPs participating in DR. VPP$_i$'s strategy set is $\mathbf{\bar{X}}=\{\Delta \lambda_{ij,t},\Delta p_{ij,t},\bar{p}_{i,t} ^{\mathrm{b}},s_{i,t}, q_{i,t} ^{\mathrm{ch}}, q_{i,t}^{\mathrm{dis}},e_{i,t}^b\}$, where the symbol $``\ \bar{\cdot} \ "$ represents parameters or decision variables whose mathematical definitions change when workload or energy migration parameters are added. Variables $\Delta \lambda_{ij,t}$ and $\Delta p_{ij,t}$ represent workload migration and energy transfer between VPP$_i$ and VPP$_j$, respectively.

System cost minimization problem (\textbf{P2}) in cooperative operation:
\begin{align}
\label{eq:coop_obj_full}
\min_{\bar{\mathbf{X}}} \quad 
\sum_{i\in \mathcal{N}} \bar{C}_i + \sum_{i\in \mathcal{N}} C_i^M - \bar{R}^{\mathrm{DR}},
\end{align}
where
\begin{align}
\label{eq:coop_obj1}
\bar{C}_i &= C_{i} ^{\mathrm{PV}}+ C_{i} ^{\mathrm{BESS}} + \sum_{t \in \mathcal{T}}\rho_{i,t}^{\mathrm{b}}\bar{p}_{i,t} ^{\mathrm{b}} + \bar{C}_{i} ^{\mathrm{QoS}},\\
\label{eq:coop_obj2}
C_i^M &= C_{i}^{\mathrm{W}}+C_{i}^{\mathrm{E}},
\end{align}

Equation \eqref{eq:coop_obj_full} minimizes the total system cost, which comprises the sum of each VPP's local operating costs $\bar{C}_i$  and migration costs , minus the total DR revenue $\bar{R}^{\mathrm{DR}}$.

To enable cooperation, we define the workload migration matrix $\mathbf{\Lambda_t} = [\Delta \lambda_{ij,t}]_{N \times N}$ and the energy sharing matrix $\mathbf{P}_t = [\Delta p_{ij,t}]_{N \times N}$ for each time slot $t$. Here, the element $\Delta \lambda_{ij,t}$ represents the amount of interactive workload transferred from VPP$_i$ to VPP$_j$, while $\Delta p_{ij,t}$ denotes the energy transferred between them. A positive value indicates a transfer from $i$ to $j$, and a negative value indicates the reverse.

These migration matrices are subject to the following constraints:
\begin{align}
\label{eq:migration_constraint}
\begin{cases}
\Delta \lambda_{ij,t}+\Delta \lambda_{ji,t}=0,\\
\Delta p_{ij,t}+\Delta p_{ji,t}=0,\\
\Delta \lambda_{ii,t}=0, \Delta p_{ii,t}=0,\\
|\Delta \lambda_{ij,t}| \leq \lambda^{\mathrm{max}}, |\Delta p_{ij,t}| \leq p^{\mathrm{max}},\\
\end{cases}
, \quad  \forall i,j, t ,
\end{align}
where the first two equations ensure conservation, the third and fourth prohibit self-migration, and the fifth and sixth impose capacity limits defined by $\lambda^{\mathrm{max}}$ and $p^{\mathrm{max}}$.

Due to workload migration, the actual computation load, server utilization ratio, DC power consumption, and electricity purchase constraint for VPP$_i$ are adjusted as follows:
\begin{align}
\label{eq:co_data_center_power}
\begin{cases}
\bar{p}_{i,t}^{\mathrm{d}} &= s_{i,t}\Bigl[e^{\mathrm{idle}} + \bigl(e^{\mathrm{peak}} - e^{\mathrm{idle}}\bigr)\bar{U}_{i,t} + (\eta - 1)e^{\mathrm{peak}}\Bigr]  \\
&\quad + e_{i,t}^b, \\
\bar{U}_{i,t} &= \frac{\bar{\lambda}_{i,t}}{s_{i,t}\,u_i}, \\
\bar{\lambda}_{i,t} &= {\lambda}_{i,t} - \sum_{j\in \mathcal{N}} \Delta \lambda_{ij,t}, 
\end{cases}
\end{align}
\begin{align}
\bar{p}_{i,t}^{\mathrm{b}} &\geq \bar{p}_{i,t}^{\mathrm{d}} - \Bigl(p_{i,t}^{\mathrm{PV}} - q_{i,t}^{\mathrm{ch}} + q_{i,t}^{\mathrm{dis}}\Bigr) + \sum_{j\in \mathcal{N}} \Delta p_{ij,t}, \forall i,j,t,
\end{align}
\begin{align}
\label{eq:coo_QOS}
\bar{C}_{i} ^{\mathrm{QoS}} = \sum_{t \in \mathcal{T}} \kappa \bar{\lambda}_{i,t}\left(\frac{1}{s_{i,t}\,u_i - \bar{\lambda}_{i,t}}\right) .
\end{align}

The migration costs are quantified using a distance matrix $D = [d_{ij}]_{N\times N}$, where $d_{ij}$ represents the distance from VPP$_i$ to VPP$_j$:
\begin{align}
C_{i}^{\mathrm{W}} &= \sum_{t \in \mathcal{T}} \omega_1 d_{ij} {( \sum_{j\in \mathcal{N}} \max(0,\Delta \lambda_{ij,t}}),\\
C_{i}^{\mathrm{E}} &= \sum_{t \in \mathcal{T}} \omega_2  d_{ij} {( \sum_{j\in \mathcal{N}} \max(0,\Delta p_{ij,t}}),
\end{align}
where $\omega_1$ and $\omega_2$ are the unit costs for workload and energy transmission, respectively.

For DR revenue calculation, the aggregate load curve is compared with the CDL:
\begin{align}
\begin{cases}
\bar{L}_{t} = \frac{\sum_{i\in \mathcal{N}} \bar{p}_{i,t} ^{\mathrm{b}} }{ \sum_{i\in \mathcal{N}} P_i^{D}}, \\[10pt]
\bar{d} = \sqrt{\sum_{t \in \mathcal{T}} \Bigl(\bar{L}_{t} - L_t ^{\mathrm{CDL}}\Bigr)^2},
\end{cases}
\end{align}
where $\bar{L}_{t}$ is the normalized aggregate load and $\bar{d}$ is the deviation from the ideal curve $L_t ^{\mathrm{CDL}}$. Defining similarity as $\varepsilon=1-\bar{d}$, the total DR revenue is $\bar{R}^{\mathrm{DR}} =\sum_{i\in \mathcal{N}} \rho_o\,\varepsilon P_i^{D}$.

\subsection{Improved Shapley Value-Based Benefit Sharing Strategy}
This section introduces a novel benefit-sharing method for VPPs in collaborative DR programs that maintains fairness principles while overcoming computational complexity through a strategic VPPO intermediary mechanism, avoiding the exponentially complex traditional Shapley value approach. 

The characteristic function $c(S)$ representing the minimum total cost achievable by any coalition $S\subseteq \mathcal{N}$ is defined as:
\begin{align}
\label{eq:union_cost}
c(S)=
\min_{\bar{\mathbf{X}}} 
\sum_{i\in \mathcal{S}} \bar{C}_i + \sum_{i\in \mathcal{S}} C_i^M -\bar{R}_{S}^{\mathrm{DR}},
\end{align}
where $\bar{C}_i$  is VPP$_i$'s operational cos, $C_i^M$ denotes the maintenance cost, and $\bar{R}_{S}^{\mathrm{DR}}$ represents the DR revenue of coalition $S$.  Operational parameters follow the model in Section III.

The three-stage  ``cost saving-commission-redistribution" framework positions VPPO as strategic intermediaries. Total collaboration savings are calculated as:
\begin{align}
V^\mathrm{save}=\sum_{i\in\mathcal{N}}c(\{i\})-c(\mathcal{N}).
\end{align}

This measure serves to establish a robust baseline for evaluating the overall value created through cooperation. To develop a sustainable coordination framework, the model allows the VPPO to extract a proportion $\gamma$ of the total saved cost as a scheduling service fee:
\begin{align}
V^{\mathrm{O}}=\gamma\cdot V_{\mathrm{save}}	.
\end{align}
where $\gamma\in[0,1)$ represents an extraction ratio parameter calibrated according to market conditions and coordination complexity.

While the standard Shapley value calculation is defined as:
\begin{align}
\phi_i(c)=\sum_{S\subseteq N\setminus\{i\}}\frac{|S|!(n-|S|-1)!}{n!}[c(S\cup\{i\})-c(S)],
\end{align}

The framework introduces a computationally efficient alternative designed to preserve the proportional contribution principle while reducing complexity from $O(2^N)$ to $O(N)$:
\begin{equation}
    \phi'_i(c)=c(\{i\})+\frac{c(\{i\})}{\sum_{j\in N}c(\{j\})} [c(N)-\sum_{j\in N}c(\{j\})].
\end{equation}

To enhance analytical clarity, the component of cost saved through cooperation is defined as:
\begin{align}
V_i^\mathrm{c}=c(\{i\})-\phi'_i(c)
\end{align}

For equitable distribution purposes, the saving ratio coefficient is calculated as:
\begin{align}
\theta_i=\frac{V_i^\mathrm{c}}{V^\mathrm{save}}=\frac{c(\{i\})-\phi'_i(c)}{\sum_{i\in\mathcal{N}}c(\{i\})-c(\mathcal{N})}.
\end{align}

Finally, the allocated cost for each VPP$_i$ (\textbf{P3}) is:
\begin{align}
    C_i^*=c(\{i\})-\theta_i (1-\gamma)\cdot V^\mathrm{save}.
\end{align}

\subsection{Uncertainty Modeling}
This study considers only uncertainties in interactive workload arrivals and PV generation, as batch workloads are typically planned tasks. For independently operating VPPs, the power balance relationship follows from equations \eqref{eq:data_center_power}, \eqref{eq:server_utilization}, and \eqref{eq:independent_external_load}:
\begin{align}
\label{eq:ind_constraint}
s_{i,t}\Biggl[e^{\mathrm{idle}} +\frac{\bigl(e^{\mathrm{peak}} - e^{\mathrm{idle}}\bigr)\lambda_{i,t}}{s_{i,t}u_i} + (\eta - 1)e^{\mathrm{peak}}\Biggr] \notag \\
+e_{i,t}^b- \Bigl(p_{i,t}^{\mathrm{PV}}- q_{i,t}^{\mathrm{ch}} + q_{i,t}^{\mathrm{dis}}\Bigr)-p_{i,t}^{\mathrm{b}} \leq 0, \quad \forall t.
\end{align}

In cooperative mode, according to \eqref{eq:co_data_center_power}, the power balance relationship is:
\begin{align}
\label{eq:co_constraint}
s_{i,t}\Biggl[e^{\mathrm{idle}} + \frac{\bigl(e^{\mathrm{peak}} - e^{\mathrm{idle}}\bigr)\bigl({\lambda}_{i,t} - \sum_{j\in \mathcal{N}} \Delta \lambda_{ij,t}\bigr)}{s_{i,t}u_i} \notag \\
+ (\eta - 1)e^{\mathrm{peak}}\Biggr]+e_{i,t}^b- \Bigl(p_{i,t}^{\mathrm{PV}}- q_{i,t}^{\mathrm{ch}} + q_{i,t}^{\mathrm{dis}}\Bigr) \notag \\
+ \sum_{j\in \mathcal{N}}\Delta p_{ij,t}-\bar{p}_{i,t}^{\mathrm{b}} \leq 0, \quad  \forall t.
\end{align}

To address uncertainties in PV generation and DC task arrivals, fuzzy parameters $\tilde{p}_{i,t}^{PV}$ and $\tilde{\lambda}_{i,t}$ are introduced. Equations \eqref{eq:ind_constraint} and \eqref{eq:co_constraint} are relaxed to power balance constraints with specified confidence level $\beta \in [0,1]$, as shown in \eqref{eq:relaxe_ind_constraint} and \eqref{eq:relaxe_co_constraint}:
\begin{align}
\label{eq:relaxe_ind_constraint}
\text{Cr}\Bigg\{s_{i,t}\Biggl[e^{\mathrm{idle}} +\frac{\bigl(e^{\mathrm{peak}} - e^{\mathrm{idle}}\bigr)\tilde{\lambda}_{i,t}}{s_{i,t}u_i} + (\eta - 1)e^{\mathrm{peak}}\Biggr] \notag \\
+e_{i,t}^b- \Bigl(\tilde{p}_{i,t}^{\mathrm{PV}}- q_{i,t}^{\mathrm{ch}} + q_{i,t}^{\mathrm{dis}}\Bigr)-p_{i,t}^{\mathrm{b}} \leq 0\Bigg\} \geq \beta,
\end{align}
\begin{align}
\label{eq:relaxe_co_constraint}
\text{Cr}\Bigg\{s_{i,t}\Biggl[e^{\mathrm{idle}} + \frac{\bigl(e^{\mathrm{peak}} - e^{\mathrm{idle}}\bigr)\bigl(\tilde{\lambda}_{i,t} - \sum_{j\in \mathcal{N}} \Delta \lambda_{ij,t}\bigr)}{s_{i,t}u_i} \notag \\
+ (\eta - 1)e^{\mathrm{peak}}\Biggr]+e_{i,t}^b- \Bigl(\tilde{p}_{i,t}^{\mathrm{PV}}- q_{i,t}^{\mathrm{ch}} + q_{i,t}^{\mathrm{dis}}\Bigr) \notag \\
+ \sum_{j\in \mathcal{N}}\Delta p_{ij,t}-\bar{p}_{i,t}^{\mathrm{b}}\leq 0\Bigg\} \geq \beta .
\end{align}

A trapezoidal fuzzy variable is represented by a quadruple of crisp numbers $(r_1,r_2,r_3,r_4)$, $(r_1 \leq r_2 \leq r_3 \leq r_4)$, with the membership function:
\begin{align}
\label{eq:affiliation_function}
\left.\mu\left(x\right)=\left\{
\begin{array}{cc}\frac{x-r_{1}}{r_{2}-r_{1}},&\text{if}\ r_{1}\leq x\leq r_{2},\\
1,&\text{if}\ r_{2}\leq x\leq r_{3} ,\\
\frac{x-r_{4}}{r_{3}-r_{4}},&\text{if}\  r_{3}\leq x\leq r_{4},\\
0,&\text{other}.
\end{array}\right.\right.
\end{align}

\begin{theorem}\label{thm:1}
If the function takes the following form:
\begin{align*}
g\left(x,\zeta\right) &= h_1\left(x\right)\zeta_1+h_2\left(x\right)\zeta_2  +\cdots+h_t\left(x\right)\zeta_t+h_0\left(x\right),
\end{align*}
where $\zeta_k$ is a trapezoidal fuzzy variable, $k=1,2,\cdots t, t\in R$. $r_{k1}-r_{k4}$ are trapezoidal membership parameters. When $\beta\geq 1/2$, the crisp equivalent of
$\text{Cr}\left\{g\left(x,\varepsilon\right)\leq0\right\}\geq\beta$ is:
\begin{align*}
&\left\{
\begin{array}{l}
h_0\left(x\right)+(2-2\beta)\sum_{k=1}^t\left[r_{k3}h_k^+\left(x\right)-r_{k2}h_k^-\left(x\right)\right] \\
\quad +(2\beta-1)\sum_{k=1}^t\left[r_{k4}h_k^+\left(x\right)-r_{k1}h_k^-\left(x\right)\right]\leq0,
\end{array}
\right.
\end{align*}
where
\begin{align*}
\begin{cases}
h_k^+\left(x\right) &= h_k\left(x\right)\vee0,\\
h_k^-\left(x\right) &= -h_k\left(x\right)\wedge0.
\end{cases}
\end{align*}
\end{theorem}

Assume that the fuzzy parameters $\tilde{p}_{i,t}^{PV}$ and $\tilde{\lambda}_{i,t}$ follow triangular fuzzy distributions, which can be represented as triples:
\begin{align}
\begin{cases}
\tilde{\lambda}_{i,t} &= (a_{i,t}^{\lambda}, b_{i,t}^{\lambda}, c_{i,t}^{\lambda}), \\
\tilde{p}_{i,t}^{PV} &= (a_{i,t}^{p}, b_{i,t}^{p}, c_{i,t}^{p}),
\end{cases}
\end{align}
where triangular fuzzy distribution is a special case of trapezoidal fuzzy distribution; when $r_2=r_3$, the trapezoid degenerates to a triangle. For triangular fuzzy variables, the parameter correspondence is: $r_1=a,r_2=r_3=b,r_4=c$.

The detailed derivation of the deterministic equivalent constraint can be found in Appendix.
From Theorem \ref{thm:1}, we get the equivalent constraint for independent operation:
\begin{align}
\label{eq:ind_equivalent constraint}
&s_{i,t}[e^{\mathrm{idle}} + (\eta - 1)e^{\mathrm{peak}}] + (q_{i,t}^{\mathrm{ch}} - q_{i,t}^{\mathrm{dis}}) \notag \\
&- p_{i,t}^{\mathrm{b}} + \frac{e^{\mathrm{peak}} - e^{\mathrm{idle}}}{u_i}[(2-2\beta)b_{i,t}^{\lambda}+ (2\beta-1)c_{i,t}^{\lambda}] \notag \\
& +e_{i,t}^b- [(2-2\beta)b_{i,t}^{p} + (2\beta-1)a_{i,t}^{p}] \leq 0 , \quad \forall t.
\end{align}

The constraint $s_{i,t}u_i \geq \lambda_{i,t}$ can be equivalently transformed into:
\begin{align}
\label{eq:s_constraint}
s_{i,t}u_i\geq (2-2\beta)b_{i,t}^\lambda+(2\beta-1)c_{i,t}^\lambda, \quad \forall t.
\end{align}

Equations \eqref{eq:qos_cost} and \eqref{eq:server_constraint} can be equivalently transformed into:
\begin{align}
\label{eq:transformed}
\begin{cases}
U_{i,t} &= \frac{(2-2\beta)b_{i,t}^\lambda+(2\beta-1)c_{i,t}^\lambda}{s_{i,t} u_i},\\
C_{i}^{\mathrm{QoS}} &= \sum_{t \in \mathcal{T}} \kappa \lambda_{i,t} \left(\frac{1}{s_{i,t} u_i - [(2-2\beta)b_{i,t}^{\lambda} + (2\beta-1)a_{i,t}^{\lambda}]}\right).
\end{cases}
\end{align}

The problem \textbf{P1} can be expressed as:
\begin{align}
\begin{cases}
\underset{\mathbf{X}}{\min} \quad 
C_{i} - R_i ^{\mathrm{DR}} \\
\text{s.t.} \quad 
\eqref{eq:SoC},\eqref{eq:SoC_constraints}, \eqref{eq:ind_equivalent constraint},\eqref{eq:s_constraint}, \eqref{eq:transformed},\\
\sum_{t \in \mathcal{T}} e_{i,t}^b = \hat{e}_{i}^b,
s_{i,t} \leq s^{\mathrm{max}},
p_{i,t}^{\mathrm{b}} \geq 0, \quad \forall t,\\
\beta \geq 1/2,
\end{cases}
\end{align}
where $\mathbf{X}=\{s_{i,t}, p_{i,t}^{\mathrm{b}}, q_{i,t}^{\mathrm{ch}}, q_{i,t}^{\mathrm{dis}},e_{i,t}^b\}$ represents the set of decision variables for the independent operations of VPP$_i$.

The crisp equivalent constraint for the cooperative case is:
\begin{align}
\label{eq:co_equivalent constraint}
&s_{i,t}[e^{\mathrm{idle}} + (\eta - 1)e^{\mathrm{peak}}] + (q_{i,t}^{\mathrm{ch}} - q_{i,t}^{\mathrm{dis}}) \notag \\
&+ \sum\limits_{j\in \mathcal{N}}\Delta p_{ij,t} - \bar{p}_{i,t}^{\mathrm{b}} - \frac{e^{\mathrm{peak}} - e^{\mathrm{idle}}}{u_i}\sum\limits_{j\in \mathcal{N}} \Delta \lambda_{ij,t} \notag \\
&+ \frac{e^{\mathrm{peak}} - e^{\mathrm{idle}}}{u_i}[(2-2\beta)b_{i,t}^{\lambda} + (2\beta-1)c_{i,t}^{\lambda}] +e_{i,t}^b \notag \\
&- [(2-2\beta)b_{i,t}^{p} + (2\beta-1)a_{i,t}^{p}] \leq 0,\quad \forall i,t,
\end{align}

\begin{align}
\label{eq:s_constraint2}
\begin{cases}
\begin{aligned}
s_{i,t}u_i &\geq (2-2\beta)b_{i,t}^\lambda+(2\beta-1)c_{i,t}^\lambda \\
&\quad - \sum_{j\in \mathcal{N}} \Delta \lambda_{ij,t}, \\
\bar{\lambda}_{i,t} &= (2-2\beta)b_{i,t}^\lambda+(2\beta-1)c_{i,t}^\lambda \\
&\quad - \sum_{j\in \mathcal{N}} \Delta \lambda_{ij,t},
\end{aligned}
\end{cases}
\quad, \forall i,j,t.
\end{align}
The problem \textbf{P2} can be expressed as:
\begin{align}
\label{eq:problem_P2}
\begin{cases}
\underset{\mathbf{\bar{X}}}{\min} \quad
\sum_{i\in \mathcal{N}} \bar{C}_i + \sum_{i\in \mathcal{N}} C_i^M - \bar{R}^{\mathrm{DR}} \\
\text{s.t.} \quad  
\eqref{eq:SoC}, \eqref{eq:SoC_constraints}, \eqref{eq:migration_constraint}, \eqref{eq:co_equivalent constraint}, \eqref{eq:s_constraint2}\\
\sum_{t \in \mathcal{T}} e_{i,t}^b = \hat{e}_{i}^b,
s_{i,t} \leq s^{\mathrm{max}}, 
\bar{p}_{i,t}^{\mathrm{b}} \geq 0, \quad \forall i,t,\\
 \beta \geq 1/2,
\end{cases}
\end{align}
where $\mathbf{\bar{X}}=\{\Delta \lambda_{ij,t},\Delta p_{ij,t},\bar{p}_{i,t} ^{\mathrm{b}},s_{i,t}, q_{i,t} ^{\mathrm{ch}}, q_{i,t}^{\mathrm{dis}},e_{i,t}^b\}$ represents the set of all decision variables when VPPs cooperate.

\section{Solution Methods}
\subsection{Solution Method for Independent Operation Scenario}
The optimization in equation \eqref{eq:independent_cost} faces challenges: integer variables $s_{i,t}$, nonlinear terms in $C_{i}^{\mathrm{QoS}}$ and $R_i^{\mathrm{DR}}$, and fuzzy uncertainty constraints.
\subsubsection{Integer Variable Relaxation and Correction}
For integer variable $s_{i,t}$, we use a relaxation-and-correction approach: relax $s_{i,t}$ to continuous, solve for optimal $s_{i,t}^*$, then apply $\hat{s}_{i,t} = \lceil s_{i,t}^* \rceil$, ensuring service quality constraint satisfaction:
\begin{align}
\hat{s}_{i,t}u_i > s_{i,t}^*u_i > \lambda_{i,t}.
\end{align}

Performing a first-order Taylor expansion of the objective function $f$ at the continuous optimal solution $s^*$:
\begin{align}
f(\hat{s}) \approx f(s^*) + \nabla f(s^*)^T(\hat{s} - s^*),
\end{align}
since $s^*$ is the optimal solution of the continuous problem, for any feasible direction $d$, we have $\nabla f(s^*)^Td \geq 0$.

Given $\hat{s} - s^* \geq 0$, the objective function error satisfies:
\begin{align}
f(\hat{s}) - f(s^*) \leq \sum_{i,t} \frac{\partial f}{\partial s_{i,t}}(s^*) \cdot (\hat{s}_{i,t} - s_{i,t}^*),
\end{align}
since $\hat{s}_{i,t} - s_{i,t}^* = 1 - (s_{i,t}^* - \lfloor s_{i,t}^* \rfloor)$, we derive:
\begin{align}
\Delta \leq \sum_{i,t} \frac{\partial f}{\partial s_{i,t}}(s^*) \cdot (1 - (s_{i,t}^* - \lfloor s_{i,t}^* \rfloor)).
\end{align}

In practical applications, since $0 \leq s_{i,t}^* - \lfloor s_{i,t}^* \rfloor < 1$ and server counts are typically large (often exceeding 10,000 units), the sensitivity of the objective function to $s_{i,t}$ is generally low, making the rounding error controllable.
\subsubsection{Nonlinear Term Handling}
The queuing delay cost is a nonlinear function. By introducing an auxiliary variable $z_{i,t}$, we transform the cost function into:
\begin{align}
C_{i}^{\mathrm{QoS}} = \sum_{t \in \mathcal{T}} \kappa\lambda_{i,t}z_{i,t}^2,
\end{align}
\begin{align}
\label{eq:auxiliary_variable1}
z_{i,t} \geq \frac{1}{s_{i,t}u_i - \lambda_{i,t}} \Rightarrow z_{i,t}(s_{i,t}u_i - \lambda_{i,t}) \geq 1 .
\end{align}
According to Theorem \ref{thm:1}, the equivalent form of equation \eqref{eq:auxiliary_variable1} is:
\begin{align}
\label{eq:auxiliary_variable2}
z_{i,t}\left[s_{i,t}u_i - (2-2\beta)b_{i,t}^\lambda + (2\beta-1)c_{i,t}^\lambda\right] \geq 1 .
\end{align}

The constraint \eqref{eq:auxiliary_variable2} is converted into a second-order cone constraint:
\begin{align}
    &\Biggl\|\begin{pmatrix} 
    2 \\ 
    z_{i,t} - \left[s_{i,t}u_i - \left((2-2\beta)b_{i,t}^\lambda + (2\beta-1)c_{i,t}^\lambda\right)\right]
    \end{pmatrix}\Biggr\|_2 \notag \\
    &\le z_{i,t} + \left[s_{i,t}u_i - \left((2-2\beta)b_{i,t}^\lambda + (2\beta-1)c_{i,t}^\lambda\right)\right]
\end{align}
The DR cost is also a nonlinear function. Let $\mathbf{p}_i^{\mathrm{b}} := [p_{i,1}^{\mathrm{b}}, p_{i,2}^{\mathrm{b}}, \cdots, p_{i,T}^{\mathrm{b}}]$ and $\mathbf{L}^{\mathrm{CDL}} := [L_1^{\mathrm{CDL}}, L_2^{\mathrm{CDL}}, \cdots, L_T^{\mathrm{CDL}}]$, then equation \eqref{eq:CDL_constraint} can be relaxed as:
\begin{align}
\left\|\frac{1}{P_i^D} \mathbf{p}_i^{\mathrm{b}} - \mathbf{L}^{\mathrm{CDL}}\right\|_2 \leq d_i .
\end{align}

The nonlinear terms are transformed into SOCP form for computational tractability.

After the aforementioned transformations, the independent operation problem (\textbf{P1}) is converted into an SOCP problem. 
\subsection{ADMM Decomposition Solution Method under Cooperative Operation}
\subsubsection{Distributed Solution Framework}

To solve the large-scale optimization problem \eqref{eq:problem_P2}, we propose a distributed ADMM-based solution framework offering parallel computation, robust convergence for non-strongly convex objectives, and efficient handling of coupled constraints. The cooperative mode exhibits coupling through: (1) global DR revenue dependent on aggregated VPP purchased power, (2) inter-VPP energy and workload migration costs, and (3) anti-symmetric migration matrix constraints. Our approach implements variable splitting to separate local decision-making from global coordination. For each VPP $i$, we define local variables $\{s_{i,t}, q_{i,t}^{\mathrm{ch}}, q_{i,t}^{\mathrm{dis}}, e_{i,t}^b, \Delta \lambda_{ij,t}^i, \Delta p_{ij,t}^i, \bar{p}_{i,t}^{b,i}\}$ and introduce global coordination variables $\{\Delta \lambda_{ij,t}^g, \Delta p_{ij,t}^g, \bar{p}_{i,t}^{b,g}\}$ with corresponding consistency constraints.

The nonlinear DR revenue component is addressed by transforming it into a second-order cone (SOC) formulation using an auxiliary variable $d$:
\begin{align}
\max_{\bar{\boldsymbol{X}}_g, d} \quad &\sum_{i\in \mathcal{N}} \rho_o (1-d) P_i^{D} \\
\text{s.t.} \quad & \left\|\bar{\boldsymbol{L}} - \boldsymbol{L}^{\mathrm{CDL}}\right\|_2 \leq d,
\end{align}
where $\bar{\boldsymbol{L}}$ is the aggregate load curve vector derived from the global variables $\bar{p}_{i,t}^{b,g}$, and $\boldsymbol{L}^{\mathrm{CDL}}$ is the target curve vector.

Based on this variable splitting approach, we construct the augmented Lagrangian function:
\begin{align*}
\mathcal{L}_{\rho}(&\bar{\boldsymbol{X}}_i, \bar{\boldsymbol{X}}_g, d, \boldsymbol{y}) = \sum_{i\in \mathcal{N}} \bar{C}_i(\bar{\boldsymbol{X}}_i) + \sum_{i\in \mathcal{N}} C_i^M(\Delta \lambda_{ij,t}^i, \Delta p_{ij,t}^i) \notag \\
&- \sum_{i\in \mathcal{N}} \rho_o (1-d) P_i^{D} + \sum_{i,j,t} y_{ij,t}^{\lambda}(\Delta \lambda_{ij,t}^i - \Delta \lambda_{ij,t}^g) \notag \\
&+ \sum_{i,j,t} y_{ij,t}^{p}(\Delta p_{ij,t}^i - \Delta p_{ij,t}^g) + \sum_{i,t} y_{i,t}^{b}(\bar{p}_{i,t}^{b,i} - \bar{p}_{i,t}^{b,g}) \notag \\
&+ \frac{\rho}{2}\sum_{i,j,t} \|\Delta \lambda_{ij,t}^i - \Delta \lambda_{ij,t}^g\|_2^2 + \frac{\rho}{2}\sum_{i,j,t} \|\Delta p_{ij,t}^i - \Delta p_{ij,t}^g\|_2^2 \notag \\
&+ \frac{\rho}{2}\sum_{i,t} \|\bar{p}_{i,t}^{b,i} - \bar{p}_{i,t}^{b,g}\|_2^2,
\end{align*}
where $y_{ij,t}^{\lambda}$, $y_{ij,t}^{p}$, and $y_{i,t}^{b}$ are Lagrangian multipliers for the consistency constraints, and $\rho > 0$ is the penalty parameter balancing primal and dual residuals.

The ADMM solution process iteratively executes three core steps in each iteration $k$. First, each VPP solves its local optimization subproblem independently:
\begin{align*}
\bar{\boldsymbol{X}}_i^{k+1} = \underset{\bar{\boldsymbol{X}}_i}{\arg\min} &\biggl\{ \bar{C}_i(\bar{\boldsymbol{X}}_i) + C_i^M(\Delta \lambda_{ij,t}^i, \Delta p_{ij,t}^i) \notag \\
&+ \sum_{j,t} y_{ij,t}^{\lambda,k}(\Delta \lambda_{ij,t}^i - \Delta \lambda_{ij,t}^{g,k}) \notag \\
&+ \sum_{j,t} y_{ij,t}^{p,k}(\Delta p_{ij,t}^i - \Delta p_{ij,t}^{g,k}) \notag \\
&+ \sum_{t} y_{i,t}^{b,k}(\bar{p}_{i,t}^{b,i} - \bar{p}_{i,t}^{b,g,k}) \notag \\
&+ \frac{\rho}{2}\sum_{j,t} \|\Delta \lambda_{ij,t}^i - \Delta \lambda_{ij,t}^{g,k}\|_2^2 \notag \\
&+ \frac{\rho}{2}\sum_{j,t} \|\Delta p_{ij,t}^i - \Delta p_{ij,t}^{g,k}\|_2^2 \notag \\
&+ \frac{\rho}{2}\sum_{t} \|\bar{p}_{i,t}^{b,i} - \bar{p}_{i,t}^{b,g,k}\|_2^2 \biggr\},
\end{align*}
subject to local constraints \eqref{eq:SoC}, \eqref{eq:SoC_constraints}, \eqref{eq:co_equivalent constraint}, and \eqref{eq:s_constraint2}.

Second, the global coordination variables are updated through two distinct mechanisms. For migration variables, we apply closed-form solutions that enforce anti-symmetric properties:
\begin{align*}
\begin{cases}
\Delta \lambda_{ij,t}^{g,k+1} &= \Pi_{\lambda^{\mathrm{max}}}\Bigl(\frac{\Delta \lambda_{ij,t}^{i,k+1} - \Delta \lambda_{ji,t}^{j,k+1}}{2} \\
&\quad + \frac{y_{ij,t}^{\lambda,k} - y_{ji,t}^{\lambda,k}}{2\rho}\Bigr) ,\\
\Delta p_{ij,t}^{g,k+1} &= \Pi_{p^{\mathrm{max}}}\Bigl(\frac{\Delta p_{ij,t}^{i,k+1} - \Delta p_{ji,t}^{j,k+1}}{2} \\
&\quad + \frac{y_{ij,t}^{p,k} - y_{ji,t}^{p,k}}{2\rho}\Bigr),
\end{cases}
\end{align*}
where $\Pi_c(x) = \min(\max(x,-c),c)$ is the projection operator ensuring capacity constraint satisfaction. For load curve coordination, the global purchased power and auxiliary variables are updated by solving:
\begin{align*}
(\bar{p}_{i,t}^{b,g,k+1}, d^{k+1}) = \underset{\bar{\boldsymbol{p}}^{b,g}, d}{\arg\min} &\biggl\{ -\sum_{i\in \mathcal{N}} \rho_o (1-d) P_i^{D} \notag \\
&+ \sum_{i,t} y_{i,t}^{b,k}(\bar{p}_{i,t}^{b,i,k+1} - \bar{p}_{i,t}^{b,g}) \notag \\
&+ \frac{\rho}{2}\sum_{i,t} \|\bar{p}_{i,t}^{b,i,k+1} - \bar{p}_{i,t}^{b,g}\|_2^2 \biggr\}
\end{align*}
subject to:
\begin{align*}
\begin{cases}
\left\|\bar{\boldsymbol{L}} - \boldsymbol{L}^{\mathrm{CDL}}\right\|_2 \leq d,\\[10pt]
\bar{L}_{t} = \frac{\sum_{i\in \mathcal{N}} \bar{p}_{i,t}^{b,g}}{\sum_{i\in \mathcal{N}} P_i^{D}},\end{cases}
 \quad \forall i,t .	
\end{align*}

Third, the Lagrangian multipliers are updated according to consistency constraint violations:
\begin{align*}
\begin{cases}
y_{ij,t}^{\lambda,k+1} &= y_{ij,t}^{\lambda,k} + \rho(\Delta \lambda_{ij,t}^{i,k+1} - \Delta \lambda_{ij,t}^{g,k+1}) ,\\
y_{ij,t}^{p,k+1} &= y_{ij,t}^{p,k} + \rho(\Delta p_{ij,t}^{i,k+1} - \Delta p_{ij,t}^{g,k+1}) ,\\
y_{i,t}^{b,k+1} &= y_{i,t}^{b,k} + \rho(\bar{p}_{i,t}^{b,i,k+1} - \bar{p}_{i,t}^{b,g,k+1}).
\end{cases}
\end{align*}

\subsubsection{Computational Framework and Complexity}

\begin{algorithm}[H]
\caption{ADMM-Based Distributed Optimization and Profit Allocation}
\label{alg:cooperative}
\begin{algorithmic}[1]
\STATE \textbf{Input:} System parameters, $\beta$, $\gamma$, $\rho$, $\epsilon$, $K_{max}$

\STATE \textbf{Phase 1: ADMM Optimization (P2)}
\STATE Initialize variables and set $k = 0$
\WHILE{$k < K_\text{max}$ AND ($r^{k} \geq \epsilon$ OR $s^{k} \geq \epsilon$)}
    \STATE Each VPP$_i$ (parallel): Solve local subproblems
    \STATE Update global coordination variables with antisymmetric constraints
    \STATE Update dual variables: $r^{k+1}$, $s^{k+1}$
    \STATE $k = k + 1$
\ENDWHILE
\STATE Calculate coalition cost $c(\mathcal{N})$

\STATE \textbf{Phase 2: Profit Allocation (P3)}
\STATE Solve independent problems for each VPP$_i$ to get $c(\{i\})$
\STATE Calculate total savings: $V^{\mathrm{save}} = \sum_{i\in\mathcal{N}}c(\{i\}) - c(\mathcal{N})$
\STATE Apply improved Shapley value allocation: $C_i^* = c(\{i\}) - \theta_i (1-\gamma) \cdot V^{\mathrm{save}}$

\STATE \textbf{Output:} $\bar{\boldsymbol{X}}^*$, $c(\{i\})$, $c(\mathcal{N})$, $\theta_i$, $C_i^*$
\end{algorithmic}
\end{algorithm}

The proposed framework is designed for computational scalability. In Phase 1, the ADMM algorithm exhibits a per-iteration wall-clock complexity of approximately $O((NT)^p + N^2T)$, with $p \in [2,3]$, primarily dictated by solving $N$ local SOCPs in parallel. In Phase 2, the improved Shapley value method reduces the computational burden for profit allocation from an exponential $O(2^N)$ to a linear $N+1$ number of optimization runs, ensuring the framework's applicability to large-scale systems.

The complete solution procedure is outlined in Algorithm \ref{alg:cooperative}. The first phase solves the cooperative optimization problem using the distributed ADMM, and the second phase allocates the resulting profits equitably among participants using the improved Shapley value method.

\section{Simulation Experiments and Result Analysis}
This section demonstrates the implementation principle of the proposed model and validates its feasibility through practical case studies.
\subsection{Experiment Settings}
We consider four of Google’s DCs in the United States as representative study subjects, using real electricity prices and weather data from each location \cite{engie_historical_data,renewables_ninja}.
The DC locations are: (1) Council Bluffs, IA, (2) Berkeley County, SC, (3) The Dalles, OR, (4) Lenoir, NC. The setting of distance matrix $D$ is shown in Table \ref{tab:distance_matrix}. 
\begin{table}
\caption{Straight line distance between the four geographical locations  (unit: \textnormal{km}) }
\label{tab:distance_matrix}
\centering
\setlength{\tabcolsep}{1.2em} 
\begin{tabular}{cccccc}
\toprule
Location & Index & 1 & 2 & 3 & 4 \\
\midrule
IA & 1 & 0 & 2048 & 2540 & 1764 \\
SC & 2 & 2048 & 0 & 4515 & 423 \\
OR & 3 & 2540 & 4515 & 0 & 4232 \\
NC & 4 & 1764 & 423 & 4232 & 0 \\
\bottomrule
\end{tabular}
\end{table}

\begin{figure}[!t]
\centering
\includegraphics[width=6.5cm]{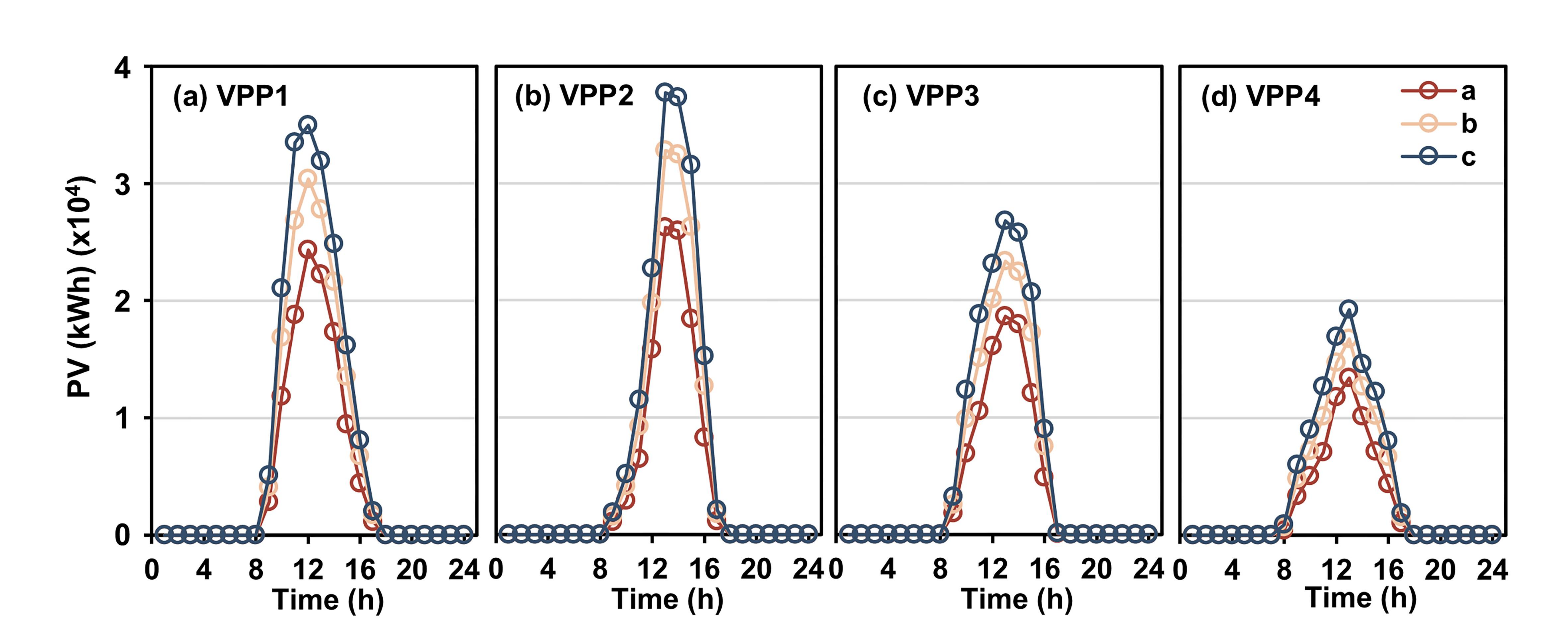}
\caption{PV power generation in different geographical locations}
\label{fig_initial_pv}
\end{figure}

\begin{figure}[!t]
\centering
\includegraphics[width=6cm]{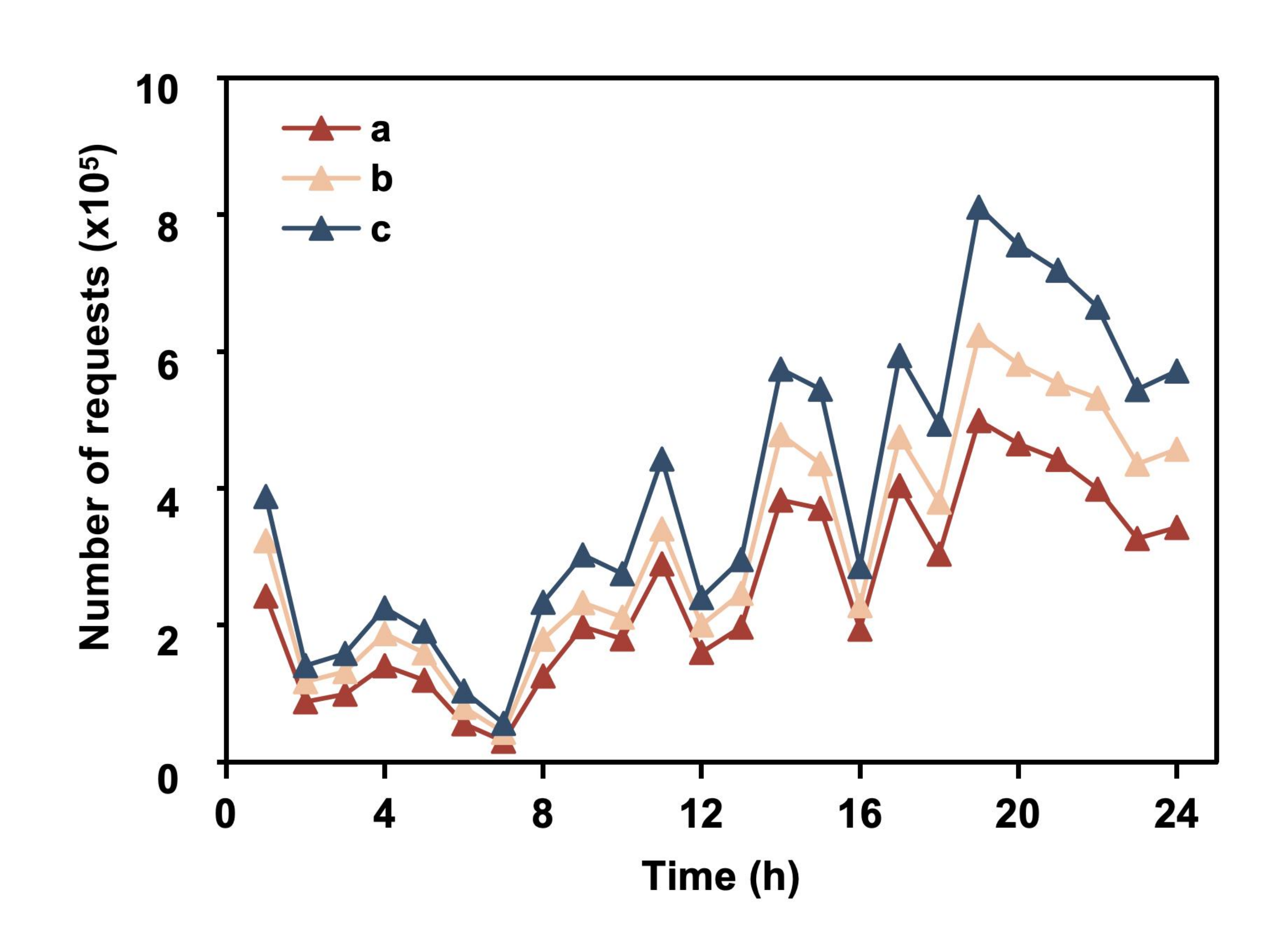}
\caption{DC interactive workload requests}
\label{fig_initial_requests}
\end{figure}

Fig. \ref{fig_initial_pv} shows the fuzzy sets of PV generation data for four geographic locations during the scheduling period. Fig. \ref{fig_initial_requests} shows the distribution of interactive workload requests during the scheduling period, it is assumed that the initial requests for the DCs are identical, where $(a,b,c)$ represent the triplets in the triangular fuzzy set.

\subsection{Result Analysis}
\begin{figure}[!t]
\centering
\includegraphics[width=7.5cm]{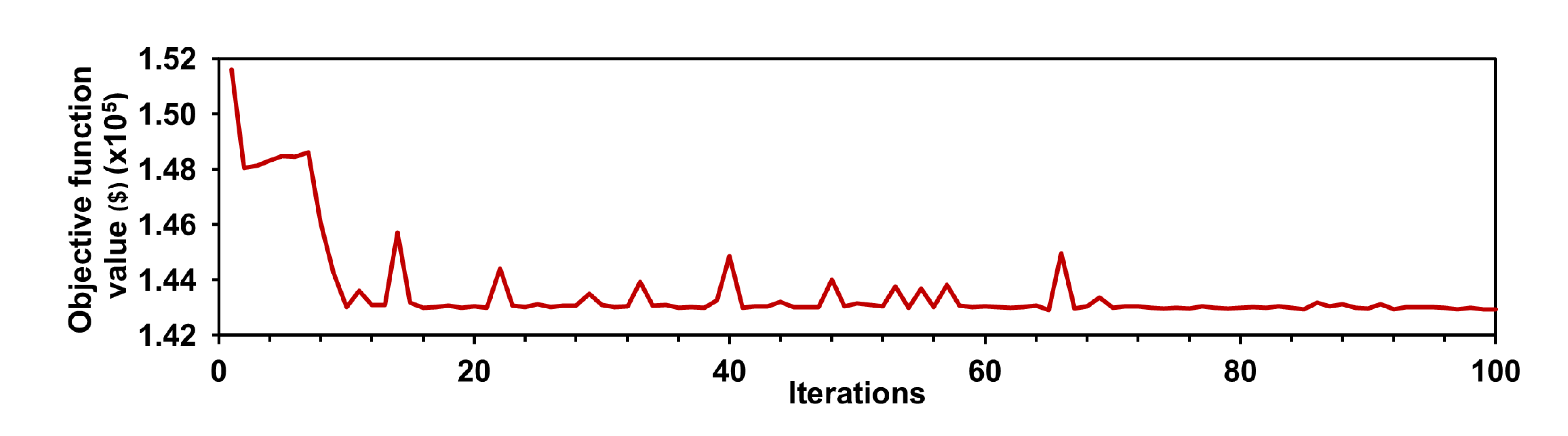}
\caption{Convergence curve of the objective function}
\label{fig_convergence_curve}
\end{figure}

\begin{figure}[!t]
\centering
\subfloat[Computational performance of the optimization algorithm]{
    \includegraphics[width=6.5cm]{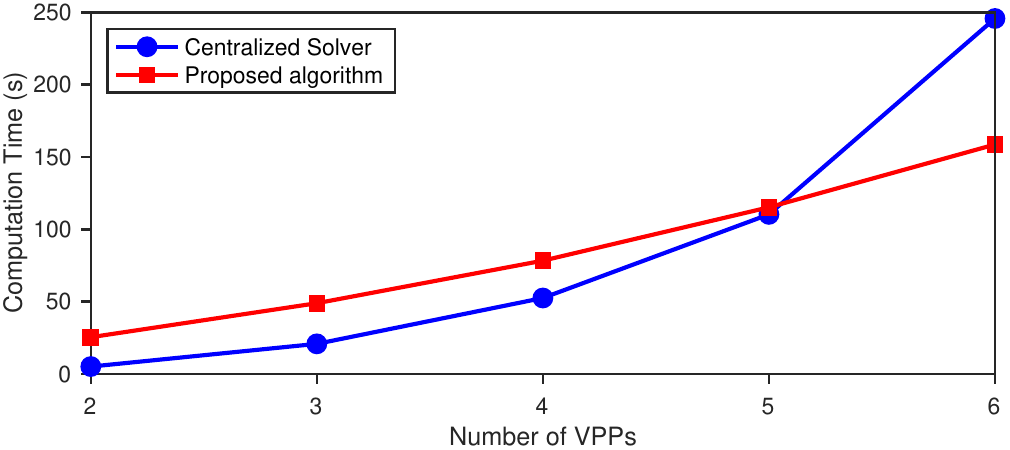}
    \label{fig:sub_opt}
}
\vspace{0.05cm} 
\subfloat[Computational performance of the benefit allocation method]{
    \includegraphics[width=6.5cm]{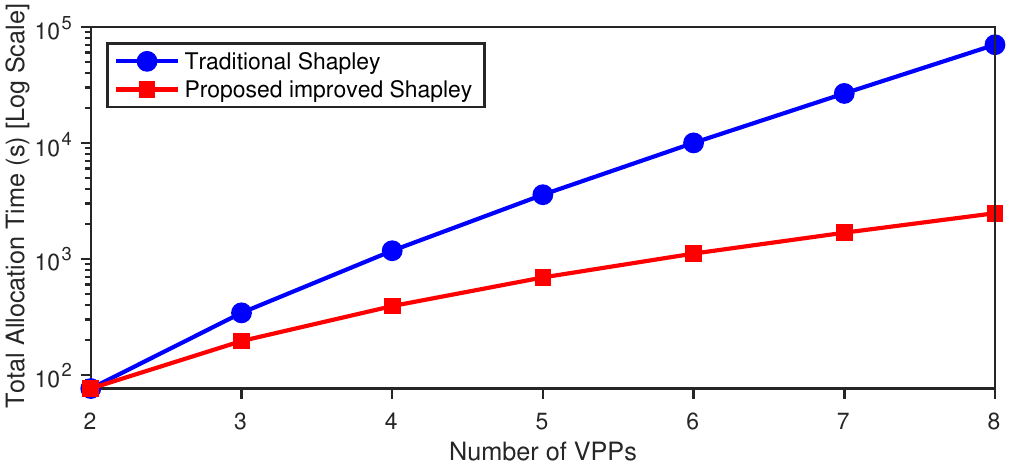}
    \label{fig:sub_alloc}
}
\caption{Scalability comparison of the proposed optimization and allocation algorithms with respect to the number of VPPs.}
\label{fig:scalability_comparison}
\end{figure}

The performance of the proposed algorithms is validated through convergence and scalability analyses, as shown in Fig.~\ref{fig_convergence_curve} and Fig.~\ref{fig:scalability_comparison}, respectively. Fig.~\ref{fig_convergence_curve} illustrates the iterative convergence of the ADMM algorithm, where the objective function exhibits a rapid initial decline and stabilizes after approximately 70 iterations. Beyond the convergence for a fixed-size problem, Fig.~\ref{fig:scalability_comparison} investigates the computational scalability of the framework as the number of VPPs ($N$) increases. As shown in Fig.~\ref{fig:scalability_comparison}\subref{fig:sub_opt}, while the centralized solver is faster for small systems, its computation time grows at a super-linear rate. In contrast, the proposed ADMM algorithm exhibits a much flatter, near-linear growth, confirming its superior scalability for the optimization phase. This advantage is even more pronounced in the benefit allocation phase, depicted in Fig.~\ref{fig:scalability_comparison}\subref{fig:sub_alloc}. The traditional Shapley value's runtime grows exponentially, quickly becoming intractable, whereas the proposed improved method's time scales linearly. Together, these results validate that the proposed framework is not only convergent but also computationally scalable, making it well-suited for practical applications in large-scale multi-VPP systems.

\begin{figure}[!t]
\centering
\includegraphics[width=6.5cm]{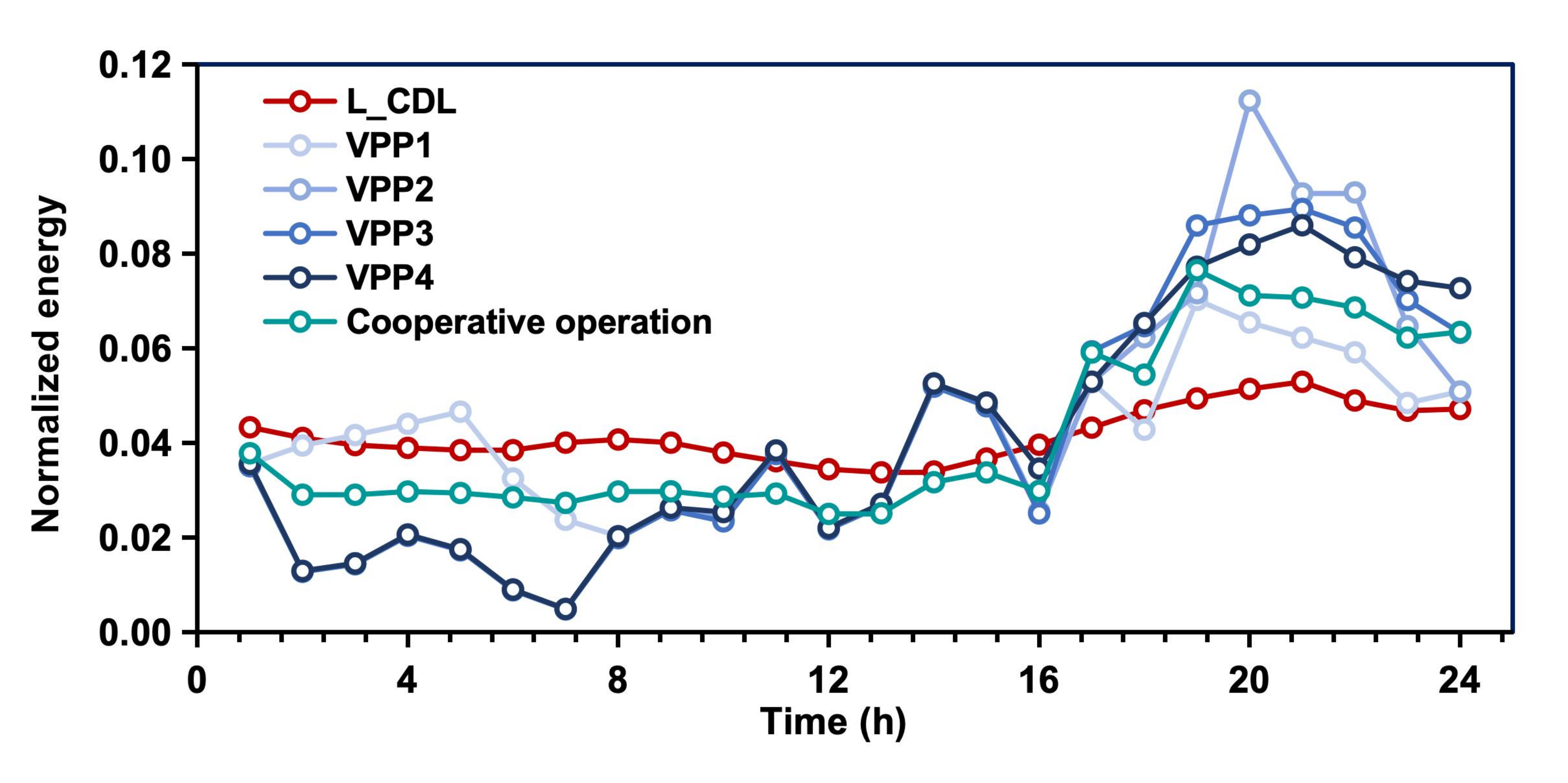}
\caption{Normalized DR curves during the scheduling period.}
\label{fig_DR_curve}
\end{figure}

Fig. \ref{fig_DR_curve} and Table \ref{tab:different_operating_incentive} demonstrate that cooperative operation significantly improves CDL tracking compared to independent operation. 
Under independent operation, VPPs show substantial deviations during peak hours (16-22h), with VPP$_2$ exhibiting the largest deviation . This corresponds with VPP$_2$'s highest Euclidean distance ($d_2=0.120$) and lowest similarity ($\varepsilon_2=0.880$) in Table \ref{tab:different_operating_incentive}. Geographic variations are evident, with VPP1 achieving the best independent performance ($d_1=0.066$, $\varepsilon_1=0.934$).
The cooperative operation profile  demonstrates superior target tracking throughout the scheduling horizon. Its Euclidean distance ($d=0.063$) represents a 47.5\% improvement over VPP$_2$ and 4.5\% over VPP$_1$, with a corresponding similarity score ($\varepsilon=0.937$) exceeding all individual performers. This enhanced precision directly translates to increased DR incentives, from 51,753.60 dollar to 53,962.79 dollar.

\begin{figure}[!t]
\centering
\includegraphics[width=6.5cm]{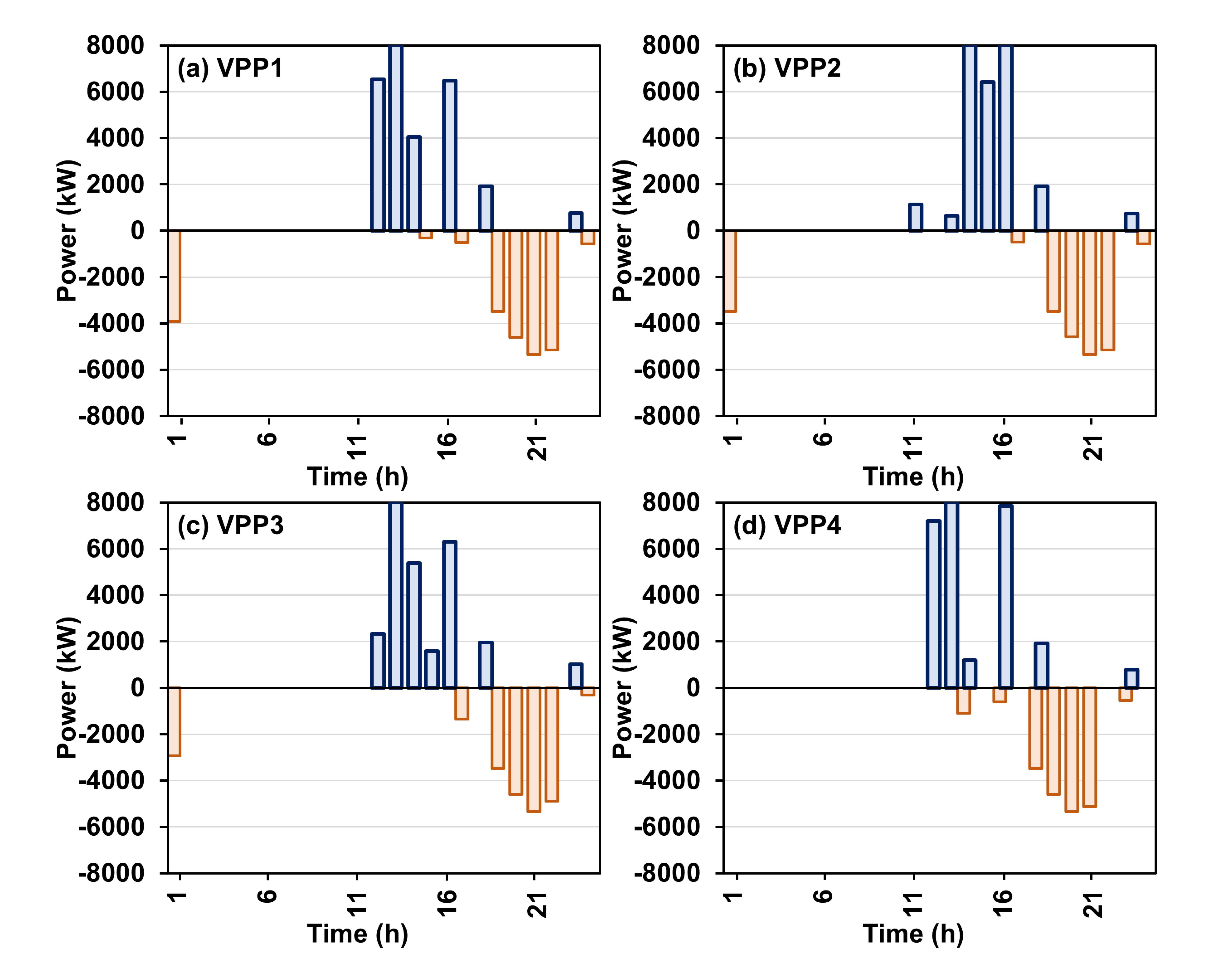}
\caption{BESS charging/discharging schedule decisions}
\label{fig_BESS}
\end{figure}

\begin{figure}[!t]
\centering
\includegraphics[width=6.5cm]{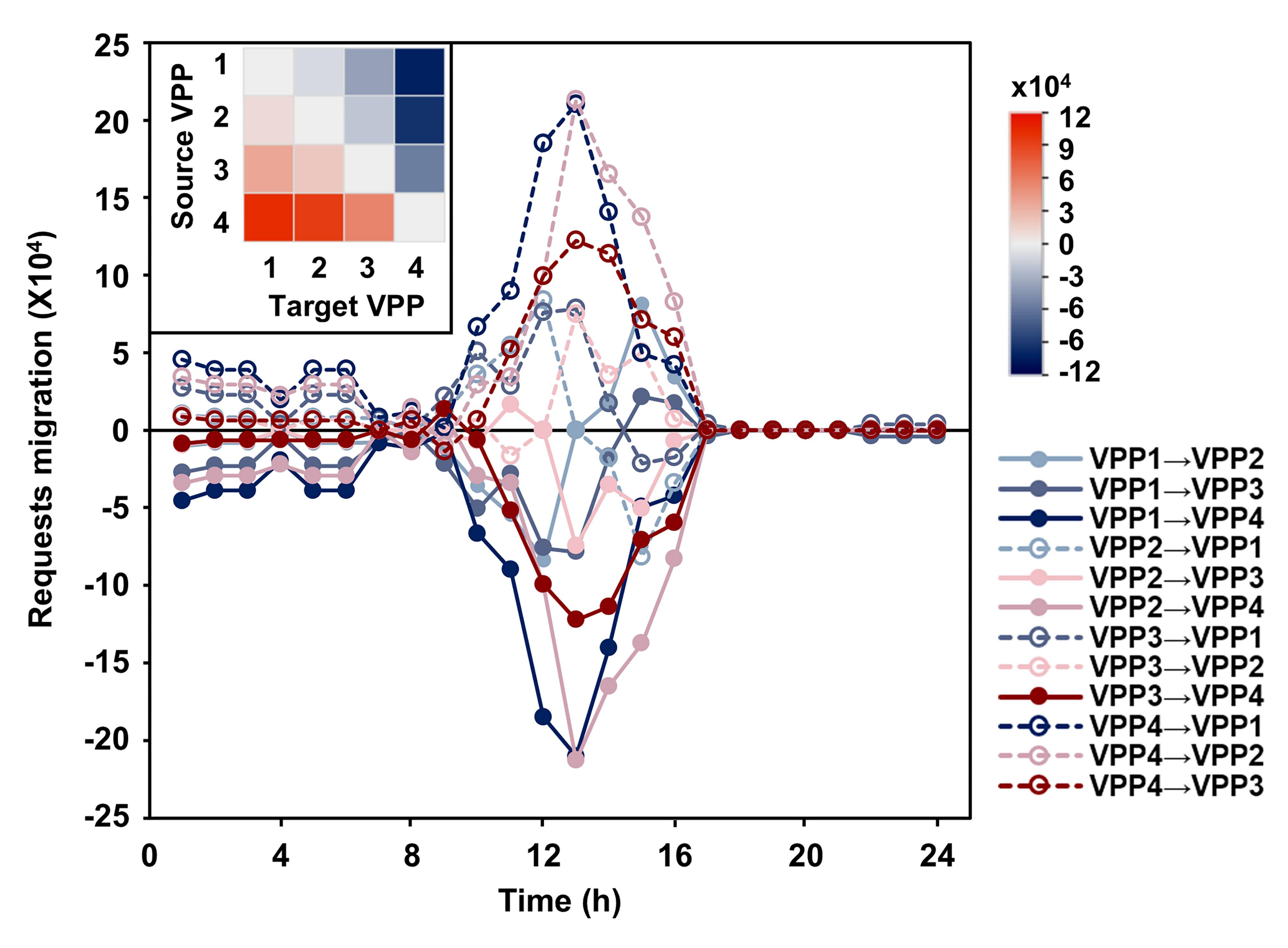}
\caption{Interactive workload migration volumes among geo-distributed VPPs}
\label{fig_workload_migration}
\end{figure}

These results validate that coordinated operation enables more precise DR through strategic exploitation of complementary resources across distributed VPPs, confirming the effectiveness of the proposed coupled migration strategy in enhancing system flexibility.
\begin{table}[htbp]
\centering
\setlength{\tabcolsep}{0.45em} 
\caption{Comparison of the effectiveness of different operating models for participating in DR}
\label{tab:different_operating_incentive}
\begin{tabular}{ccccc c}
\toprule
Mode of operation & \multicolumn{4}{c}{Independent operation} & \multirow{2}{*}{\begin{tabular}[c]{@{}c@{}}Cooperative\\operation\end{tabular}} \\
\cmidrule(lr){1-5}
DR parameters & VPP1 & VPP2 & VPP3 & VPP4 & \\
\midrule
Euclidean distance ($d_i,d$) & 0.066 & 0.120 & 0.113 & 0.107 & 0.063 \\
similarity ($\varepsilon_i,\varepsilon$) & 0.934 & 0.880 & 0.887 & 0.893 & 0.937 \\
Total DR incentive (\$) & \multicolumn{4}{c}{51753.60} & 53962.79 \\
\bottomrule
\end{tabular}
\end{table}
Fig. \ref{fig_BESS} reveals that all BESS units across the four VPPs exhibit synchronized operation patterns despite geographical separation, with charging predominantly during midday (11-16h) and discharging during evening price peaks (18-22h), show remarkable consistency, suggesting electricity price signals drive BESS dispatch decisions.  This coordinated BESS operation provides crucial temporal flexibility.
Fig. \ref{fig_workload_migration} illustrates workload migration dynamics between VPPs. The heatmap (inset) quantifies the spatial distribution of total migration throughout the scheduling period. The heatmap reveals strong complementary relationships, with VPP$_4$ functioning as the primary workload source to VPP$_1$ and VPP$_2$. Temporally, migrations concentrate during hours 10-16, with peak transfers reaching approximately $22 \times 10^4$ units around 12h. Morning hours (0-8h) exhibit modest transfers, while activity becomes minimal after 16h. 

\begin{figure}[!t]
\centering
\includegraphics[width=6.5cm]{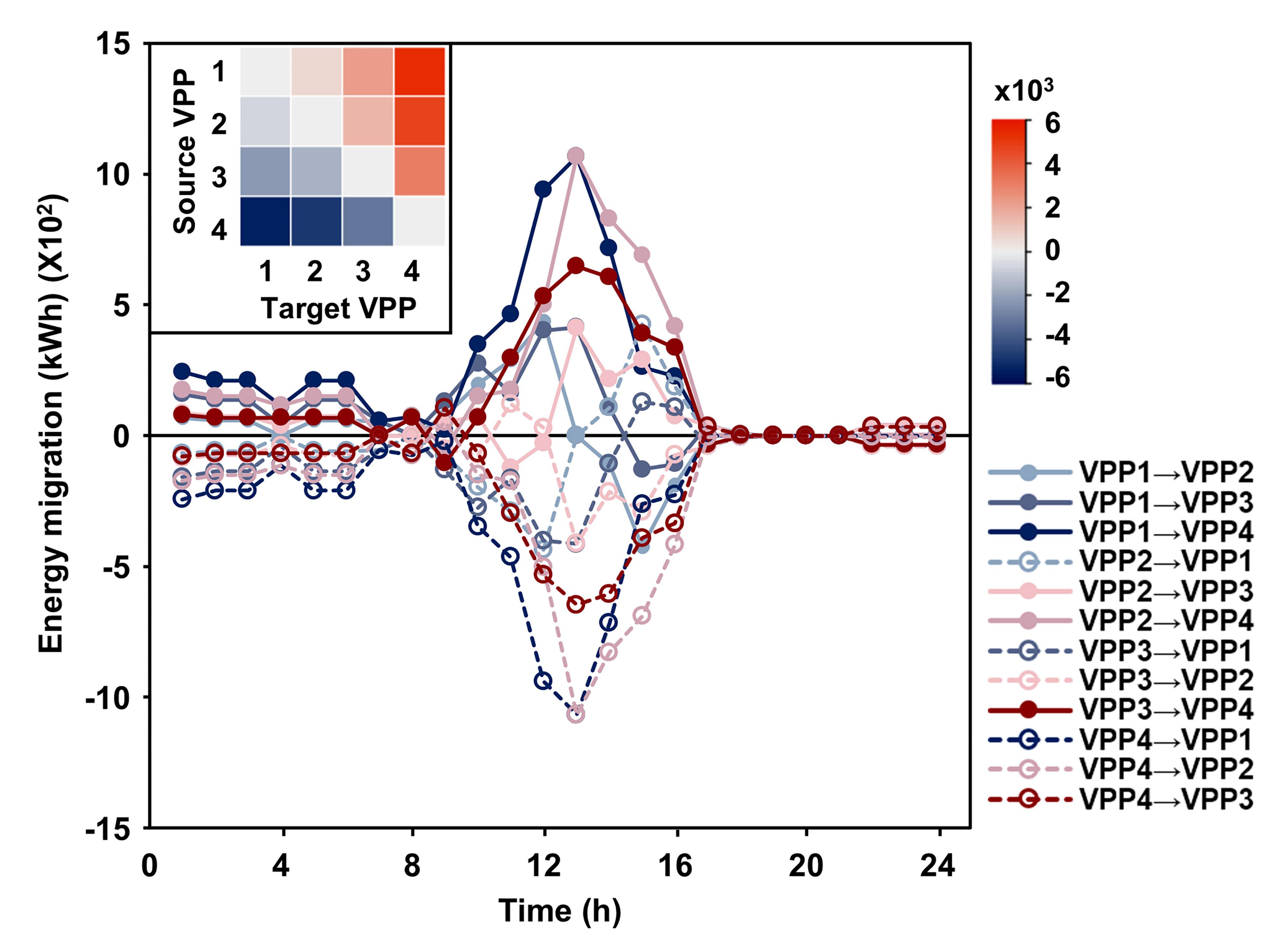}
\caption{Energy workload migration volumes among geo-distributed VPPs}
\label{fig_energy_migration}
\end{figure}

This pronounced temporal concentration aligns with peak renewable generation periods and precedes evening electricity price spikes, validating the effectiveness of the antisymmetric matrix formulation in capturing strategic bidirectional workload exchanges that maintain computational resource conservation.
Fig. \ref{fig_energy_migration} depicts bidirectional energy migration across the VPP network. The heatmap (inset) illustrates aggregate energy transfer volumes throughout the scheduling period. Spatially, VPP$_1$ to VPP$_4$ exhibits the strongest positive energy flow ($6\times 10^3$ kWh) with VPP$_4$ to VPP$_1$ showing corresponding negative flow, creating a clear diagonal antisymmetry. Temporally, energy transfers concentrate during 10-16h, with peak volumes occurring around 12-13h. Notably, the predominant energy flow direction directly opposes the primary workload migration. This inverse directional relationship validates the theoretical premise of substitutional mechanisms—energy flows to regions with high computational demand while workload shifts to areas with abundant energy resources, optimizing system-wide efficiency through complementary spatial exchanges.
Table \ref{tab:operation_costs} demonstrates the economic advantages of cooperative operation across all VPPs. Under independent operation, the total system cost reaches 156,949.0 dollar, while cooperative operation reduces this to 146,571.5 dollar, yielding significant overall savings of 10,377.5 dollar (6.6\%). Confirming that the proposed bidirectional migration strategy effectively leverages geographical complementarity to reduce system-wide operational costs while maintaining individual economic incentives for all participants.
\begin{table}
\caption{Comparison of operation costs for different VPPs}
\label{tab:operation_costs}
\centering
\setlength{\tabcolsep}{0.7em} 
\begin{tabular}{lccccc}
\toprule
\textbf{\makecell[l]{Mode of\\operation}} & \textbf{VPP1} & \textbf{VPP2} & \textbf{VPP3} & \textbf{VPP4} & \textbf{SUM} \\
\midrule
\makecell[l]{Independent\\operation (\$)} & 47432.3 & 42628.4 & 27244.2 & 39644.1 & 156949.0 \\
\makecell[l]{Cooperative\\operation (\$)} & 43611.3 & 41717.0 & 25904.5 & 35338.7 & 146571.5 \\
\bottomrule
\end{tabular}
\end{table}

\section{Conclusion}
This paper proposed an energy-workload coupled migration optimization strategy for VPPs with DCs. This paper established a coordinated migration model based on antisymmetric matrices, derived deterministic equivalent transformations of fuzzy chance constraints to address uncertainties, introduced an improved Shapley value allocation method reducing computational complexity from $O(2^N)$ to $O(N)$ compared to traditional Shapley value calculation, and designed a distributed ADMM algorithm with variable separation that effectively decomposes the globally coupled problem into N local subproblems and one coordination problem with closed-form solutions. Simulations using real data from Google's DCs validated the effectiveness of our approach in reducing operational costs while enhancing DR curve tracking precision. Future research could explore adaptive tuning of penalty parameters and migration strategies to cope with real-time system dynamics and market volatility.

\appendix

\section{Derivation of Deterministic Equivalent Constraints}

This appendix provides a detailed derivation for the deterministic equivalent of the fuzzy chance constraints used in this paper. All transformations are based on Theorem 1.

\subsection{Transformation of the Power Balance Constraint}
The primary transformation is applied to the power balance constraint for independent operation, as shown in (34). First, the constraint is standardized into the linear form $g(x, \zeta) = \sum_{k} h_k(x)\zeta_k + h_0(x) \le 0$, where $\zeta_k$ are the fuzzy variables.

The fuzzy chance constraint is:
\begin{align*}
	\text{Cr}\Bigg\{ &\underbrace{\left(\frac{e^{\mathrm{peak}} - e^{\mathrm{idle}}}{u_i}\right)}_{h_1(x)} \tilde{\lambda}_{i,t} + \underbrace{(-1)}_{h_2(x)} \tilde{p}_{i,t}^{\mathrm{PV}} \\
	&+ \underbrace{\left( s_{i,t}[e^{\mathrm{idle}} + \dots] + e_{i,t}^b + q_{i,t}^{\mathrm{ch}} - q_{i,t}^{\mathrm{dis}} - p_{i,t}^{\mathrm{b}} \right)}_{h_0(x)} \\
	&\leq 0 \Bigg\} \geq \beta
\end{align*}
The signs of the coefficients for the fuzzy variables are analyzed as follows:
\begin{enumerate}
	\item The coefficient for $\tilde{\lambda}_{i,t}$ is $h_1(x) = (e^{\mathrm{peak}} - e^{\mathrm{idle}}) / u_i$, which is always positive. Thus, $h_1^+(x) = h_1(x)$ and $h_1^-(x) = 0$.
	\item The coefficient for $\tilde{p}_{i,t}^{\mathrm{PV}}$ is $h_2(x) = -1$, which is always negative. Thus, $h_2^+(x) = 0$ and $h_2^-(x) = -h_2(x) = 1$.
\end{enumerate}
For triangular fuzzy distributions $(a, b, c)$, the parameters of the equivalent trapezoidal distribution are $r_1=a, r_2=b, r_3=b, r_4=c$. Substituting these components into the general formula from Theorem 1 yields:
\begin{align*}
	h_0(x) &+ (2-2\beta) [ (b_{i,t}^{\lambda} h_1(x) - b_{i,t}^{\lambda} \cdot 0) + (b_{i,t}^{p} \cdot 0 - b_{i,t}^{p} \cdot 1) ] \\
	&+ (2\beta-1) [ (c_{i,t}^{\lambda} h_1(x) - a_{i,t}^{\lambda} \cdot 0) + (c_{i,t}^{p} \cdot 0 - a_{i,t}^{p} \cdot 1) ] \leq 0
\end{align*}
Grouping terms by $h_1(x)$ yields:
\begin{align*}
	h_0(x) + h_1(x)\left[(2-2\beta)b_{i,t}^{\lambda} + (2\beta-1)c_{i,t}^{\lambda}\right] \\
	- \left[(2-2\beta)b_{i,t}^{p} + (2\beta-1)a_{i,t}^{p}\right] \le 0
\end{align*}
By substituting back the original expressions, we arrive at the final deterministic constraint, which is identical to (38) in the main text.

\subsection{Transformation of Other Constraints}
Other constraints involving fuzzy variables are transformed using the same principles.

\subsubsection{Server Capacity Constraint}
The server capacity constraint $s_{i,t}u_i \ge \tilde{\lambda}_{i,t}$ is written as $\text{Cr}\{ \lambda_{i,t} - s_{i,t}u_i \le 0 \} \ge \beta$. Applying Theorem 1 directly gives:
\begin{align*}
	-s_{i,t}u_i + (2-2\beta)b_{i,t}^{\lambda} + (2\beta-1)c_{i,t}^{\lambda} \le 0
\end{align*}
which is rearranged to match (39): $s_{i,t}u_i \ge (2-2\beta)b_{i,t}^{\lambda} + (2\beta-1)c_{i,t}^{\lambda}$.

\subsubsection{Cooperative Mode Constraints}
For constraints in the cooperative mode, such as (42), the additional migration terms ($\Delta\lambda_{ij,t}$, $\Delta p_{ij,t}$) are deterministic decision variables and are simply included in the $h_0(x)$ term. The transformation logic for the fuzzy variables remains identical.

\subsubsection{Corrected QoS Constraint Transformation}
The paper's original transformation of the Quality of Service (QoS) constraint in (8) and (51) contains errors. The corrected derivation is as follows. The deterministic form of the hyperbolic constraint (the corrected version of (51)) is:
\begin{align*}
	z_{i,t}\left[s_{i,t}u_i - \left((2-2\beta)b_{i,t}^\lambda + (2\beta-1)c_{i,t}^\lambda\right)\right] \geq 1
\end{align*}
This constraint is then transformed into its equivalent second-order cone form (the corrected version of (8)):
\begin{align*}
	&\Biggl\|\begin{pmatrix} 
	2 \\ 
	z_{i,t} - \left[s_{i,t}u_i - \left((2-2\beta)b_{i,t}^\lambda + (2\beta-1)c_{i,t}^\lambda\right)\right]
	\end{pmatrix}\Biggr\|_2 \\
	&\le z_{i,t} + \left[s_{i,t}u_i - \left((2-2\beta)b_{i,t}^\lambda + (2\beta-1)c_{i,t}^\lambda\right)\right]
\end{align*}

\bibliographystyle{IEEEtran}
\bibliography{refs}
\vfill
\end{document}